\documentclass[12pt]{article}
\pdfoutput=1
\usepackage{hyperref}
\usepackage{bm}
\usepackage{graphics}
\usepackage{graphicx}
\usepackage{setspace}
\usepackage[authoryear]{natbib}
\usepackage{amsmath,amsthm,amssymb}
\usepackage{titlesec} 
\usepackage{algorithm}
\usepackage{algorithmic}
\onehalfspace

\title{\vspace{-15mm}\fontsize{24pt}{10pt}\selectfont\textbf{Penalized complexity priors for degrees of freedom in Bayesian P-splines}}

\author{
\large
\textsc{Massimo Ventrucci}\thanks{Corresponding author. Email: \texttt{massimo.ventrucci@unibo.it}} and \textsc{H{\aa}vard Rue}
\\[2mm] 
\normalsize Department of Statistical Sciences, University of
Bologna, Bologna, Italy\\ 
\normalsize Department of Mathematics,
Norwegian Institute of Technology, Trondheim,
Norway
\vspace{-5mm}
}
\date{}

\begin{document}
\maketitle 

\begin{abstract}
Bayesian P-splines assume an intrinsic Gaussian Markov random field prior on the spline coefficients, conditional on a precision hyper-parameter $\tau$. Prior elicitation of $\tau$ is difficult. To overcome this issue we aim to building priors on an interpretable property of the model, indicating the complexity of the smooth function to be estimated. Following this idea, we propose Penalized Complexity (PC) priors for the number of effective degrees of freedom. We present the general ideas behind the construction of these new PC priors, describe their properties and show how to implement them in P-splines for Gaussian data.

\end{abstract}

\section{Introduction}

Penalized spline (P-spline) regression is a well established and numerically stable approach for smoothing \citep{eilers-1996,ruppert2003semiparametric}. Typically, P-spline components are implemented in Bayesian additive regression models \citep{fahrmeir-2013-book} to fit non linear covariate effects or higher dimensional effects such as spatial and spatio-temporal smooth trends. The P-spline approach proposed by \cite{eilers-1996} uses equally-spaced B-splines and constructs a smooth function as the sum of these B-splines scaled by spline coefficients. A regularizing penalty is assumed on these coefficients to control the degree of smoothness of the fitted function. A common approach is to penalize the sum of second order squared differences between adjacent spline coefficients, but specific penalties can be designed to drive the fit towards desired features \citep{eilers-2010}. This is a very useful strategy in presence of prior information about the shape, or degree of smoothness, of the function to be estimated. 

The Bayesian approach to P-spline by \cite{brezger-2004} assumes an Intrinsic Gaussian Markov Random Field (IGMRF) prior on the spline coefficients. An IGMRF is a multivariate normal distribution with rank deficient precision matrix $\bm Q(\tau)$, depending on a precision hyper-parameter $\tau$. Similarly to a regularizing penalty, the IGMRF forces the spline coefficients to be shrunk towards an infinite smooth model, which we will denote as the \textit{base model}. The degree of smoothness of the base model depends on the order of the IGMRF; for instance, an IGMRF of (polynomial) order $2$ forces shrinkage towards a linear trend, i.e. a polynomial of degree one \citep{rue-2005}.


The amount of shrinkage towards the corresponding base model depends on the IGMRF precision $\tau$. The prior $\pi(\tau)$ can have a substantial impact on the posterior distribution of the spline coefficients and hence, to some extent, on the shape of the fit. A common strategy in Bayesian P-splines is to adopt the conjugate Gamma family, i.e. $\text{Gamma}(a,b)$, with shape $a$ and rate $b$ \citep{Fahrmeir_kneibt-2009, brezger-2004}. \cite{brezger-2004} suggest to choose $a=1$ and small $b$, e.g., $b=5 \cdot 10^{-4}$, leading to a diffuse prior for $\tau^{-1}$. \cite{Jullion-2007} note that the choice of $b$ clearly affects the smoothness of the fitted curve, when sample size or signal-to-noise ratio is small, and propose a mixture of Gamma distributions with different $b$ values. Another popular choice is the $\text{Gamma}(\epsilon, \epsilon)$, with small $\epsilon$ (e.g. $\epsilon=0.001$, which is the default option in the software \textit{BayesX} \citep{bayesx}) as an attempt of vagueness on $\tau^{-1}$. The suitability of the Gamma family as a noninformative prior for the scale parameters in hierarchical models has been debated in the literature \citep{gelman-2006}; overfitting due to Gamma priors has been demonstrated in \cite{FruhwirthSchnatter-2010, FruhwirthSchnatter-2011, pcprior}. In particular, in Bayesian P-splines, the main difficulty with using a Gamma prior on $\tau$ is that $\tau$ scales differently according to the amount of noise present in the data and the number (and location) of knots selected by the user. 

The present work proposes a new prior for $\tau$ which is informative about model complexity and implicitly accounts for different choices about number (and location) of knots. A suitable measure of complexity of the P-spline model is the number of \textit{effective degrees of freedom}, in the following denoted as \textit{d}, calculated as the trace of the hat matrix \citep{hastie-1990-book}. The value $d$ relates to the degree of a polynomial equivalent to the smooth function to be estimated. An expert user who has a prior guess about the shape of this function may find easy to elicit $d$. As an example, for a monotonic cubic trend one may elicit $d$ in a range between $3$ and $5$ and assign very low prior probability to $d>5$. The key point is that, in presence of this prior information, elicitation of a range for $d$ is intuitive and immediate, whereas elicitation of a distribution for $\tau$, directly, is very difficult. 

The challenge is to design a prior distribution on a model property (i.e., $d$) rather than on a parameter of the model (i.e., $\tau$). To achieve this, we follow the Penalized Complexity (PC) prior approach proposed by \cite{pcprior}. Within this framework, we derive the PC prior for $d$ and calibrate it by two intuitive parameters: $U$, an  upper bound for $d$ and $\alpha$, the prior probability assigned to $d>U$. In the example aforementioned, the user would only need to set $U$ equal to $5$ and $\alpha=0.01$, or some other small value. As a further challenging point, note that $d$ depends on the noise variance characterizing the dataset. Thus, implementing the proposed PC prior for degrees of freedom in real datasets, where the noise variance is typically unknown, implies defining a joint prior on two quantities, the IGMRF precision and the noise precision.

The plan of the paper is as follows. In section \ref{sec:pspline}, the Bayesian P-spline approach is revised with focus on the challenges to be addressed in defining a prior for $\tau$. The principles behind the construction of a PC prior for $\tau$ are revised in section \ref{sec:pcprior-review}. In section \ref{sec:pcprior-df-pspline}, the PC prior for $d$ is derived and its properties described in the case of known noise variance. In section \ref{sec:joint-pcprior}, we show how to implement the PC prior for $d$ when the noise variance is unknown, focusing on an additive P-splines model framework. Results from a simulation study assessing the impact of the proposed prior compared to standard Gamma priors and other PC priors proposed in the recent literature are illustrated in section \ref{sec:simulation-study}. An application of these new priors in a P-spline model for nitrate concentrations observed in river \textit{Oglio}, Lombardia Region, Italy, is illustrated in section \ref{sec:realdata}. The paper closes with a discussion in section \ref{sec:discussion}.

\section{Background on P-splines}
Let $\bm y=[y_1,...,y_n]^{\textsf{T}}$ be observations of a response variable, $\bm x$ be a continuous covariate, $f$ be a smooth function describing the effect of the covariate on the response and $\bm\epsilon$ be independent errors with zero mean and variance $\tau_{\epsilon}^{-1}$. The P-spline model \citep{eilers-1996} is $y = f(\bm x) + \bm\epsilon$, $f(\bm x) = \bm{B} \bm\beta$, where $\bm B$ is a $n \text{ x } K$ basis matrix containing $K$ B-spline functions built on a set of equally-spaced (for simplicity) knots within the covariate domain, while $\bm \beta$ is a $K \text{ x } 1$ vector of unknown spline coefficients. The method requires to select a generous number of knots to over-fit the data, to then add a penalty on $\bm \beta$ which smooths adjacent spline coefficients. In the frequentist approach, $\bm \beta$ is estimated via penalized maximum likelihood, conditional on a tuning parameter regulating the degree of smoothness of $f$. The optimal tuning can be found via cross validation \citep{wood-2006} or estimated via restricted maximum likelihood in a mixed model representation (\cite{ruppert2003semiparametric}, ch. 4). P-splines are widely used in generalized additive models \citep{hastie-1990-book, wood-2006} or structured additive regression models \citep{Fahrmeir04penalizedstructured,fahrmeir-2013-book}. Higher-dimensional smooth functions can also be represented as P-splines, using the tensor product of marginal B-spline bases \citep{eilers-2006, currie-2006}. For a systematic presentation of the different approaches to penalized spline regression see \cite{ruppert2003semiparametric}; for an excellent review of spline methods and their applications in statistical modelling see \cite{hastie-element-statistical-learning}, \cite{wakefield-book} and \cite{wood-2006}.

\subsection{Bayesian P-splines}
\label{sec:pspline}
The Bayesian approach to P-splines \citep{brezger-2004} assumes an IGMRF prior on the spline coefficients, 
\begin{equation}
\pi(\bm \beta|\tau_{\beta}) = (2\pi)^{-\texttt{rank}(\bm R)/2} (|\tau_{\beta} \bm R|^{*})^{1/2} \exp\left\{-\frac{\tau_{\beta}}{2}\bm \beta^{\textsf{T}} \bm R \bm \beta\right\}
\label{eq:igmrf}
\end{equation}
where the precision $\bm Q(\tau_{\beta})$ is given by $\tau_{\beta} \bm R$. Matrix $\bm R$ is denoted as the structure of the IGMRF, i.e. a $K \text{ x } K$ sparse matrix with non-zero entries indicating conditional dependencies among the spline coefficients, $\tau_{\beta}$ is a scalar precision hyper-parameter and $|\tau_{\beta} \bm R|^{*}$ is the generalized determinant. Throughout the paper we will assume $\bm R=\bm D^{\textsf{T}}_r \bm D_r$, where $\bm D_r$ is a $(K-r) \text{ x } K$ matrix such that $\bm D_r \bm \beta=\Delta^r \bm \beta$ \citep{eilers-2006}, with $\Delta^r$ the $r^{th}$-order difference operator.
In this form, $\bm R$ is the structure of an $r^{th}$-order random walk on $\bm \beta$ (\cite{rue-2005} ch. 3) with $\text{rank}(\bm R)=K-r$, where $r$ indicates the order of the IGMRF (\ref{eq:igmrf}).

The IGMRF (\ref{eq:igmrf}) describes deviation from a base model, which is a polynomial of degree $(r - 1)$. The amount of deviation depends on $\tau_{\beta}$. A fully Bayesian specification requires priors on $\tau_{\beta}$ and $\tau_{\epsilon}$. Since we usually have enough information in the data to estimate $\tau_{\epsilon}$, the prior $\pi(\tau_{\epsilon})$ has less impact on the fit. The hyper-parameter $\tau_{\beta}$ enters at a lower level in the hierarchy, the data bring little information on it and the prior $\pi(\tau_{\beta})$ can have a substantial impact on the posterior distribution of $\bm \beta$ and, as a consequence, on the smoothness of $f$. Therefore, specification of $\pi(\tau_{\beta})$ should be as consistent as possible with the prior information actually available about the smoothness of the function to be estimated. 

\begin{figure}
\centerline{
\includegraphics[width=0.5\textwidth]{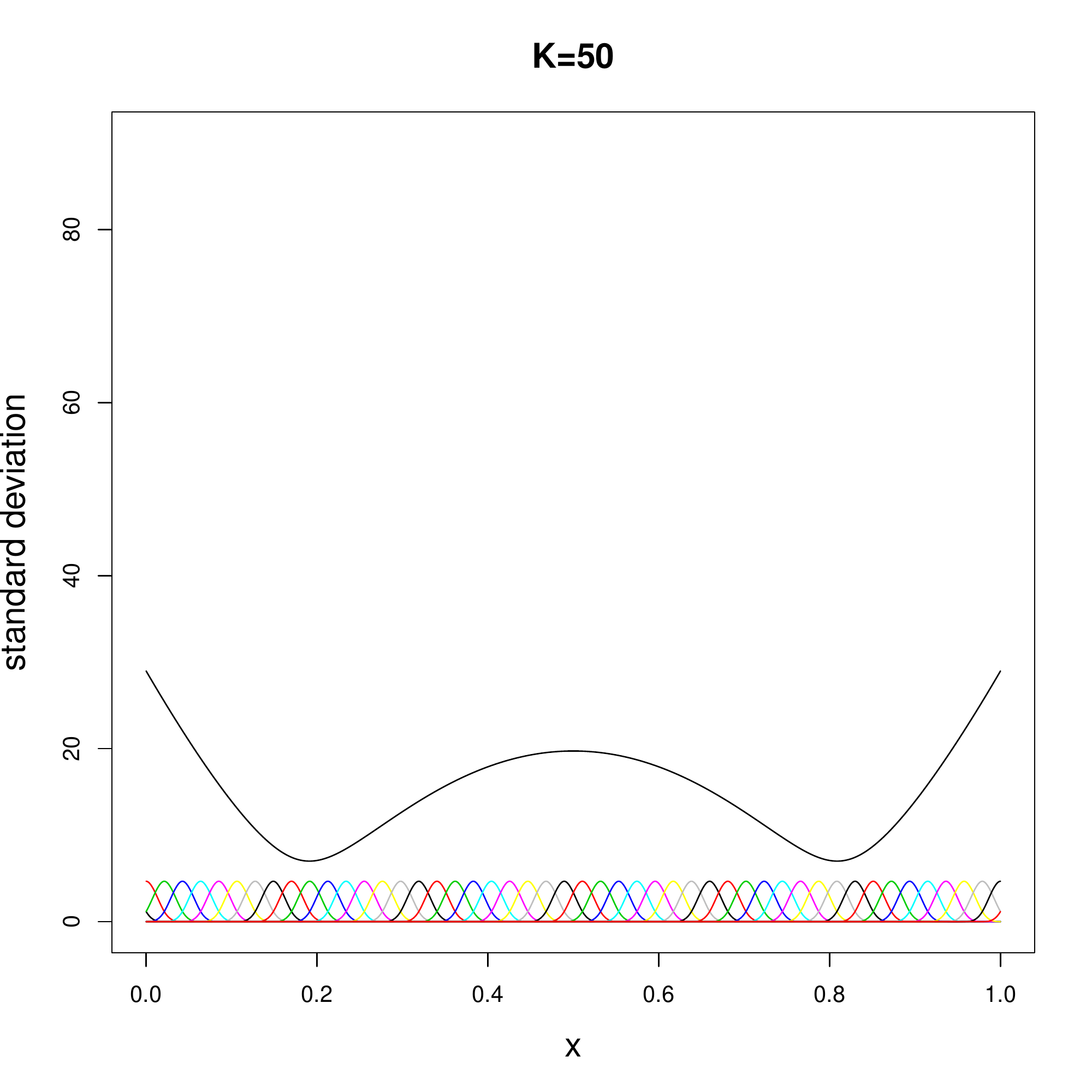}
\includegraphics[width=0.5\textwidth]{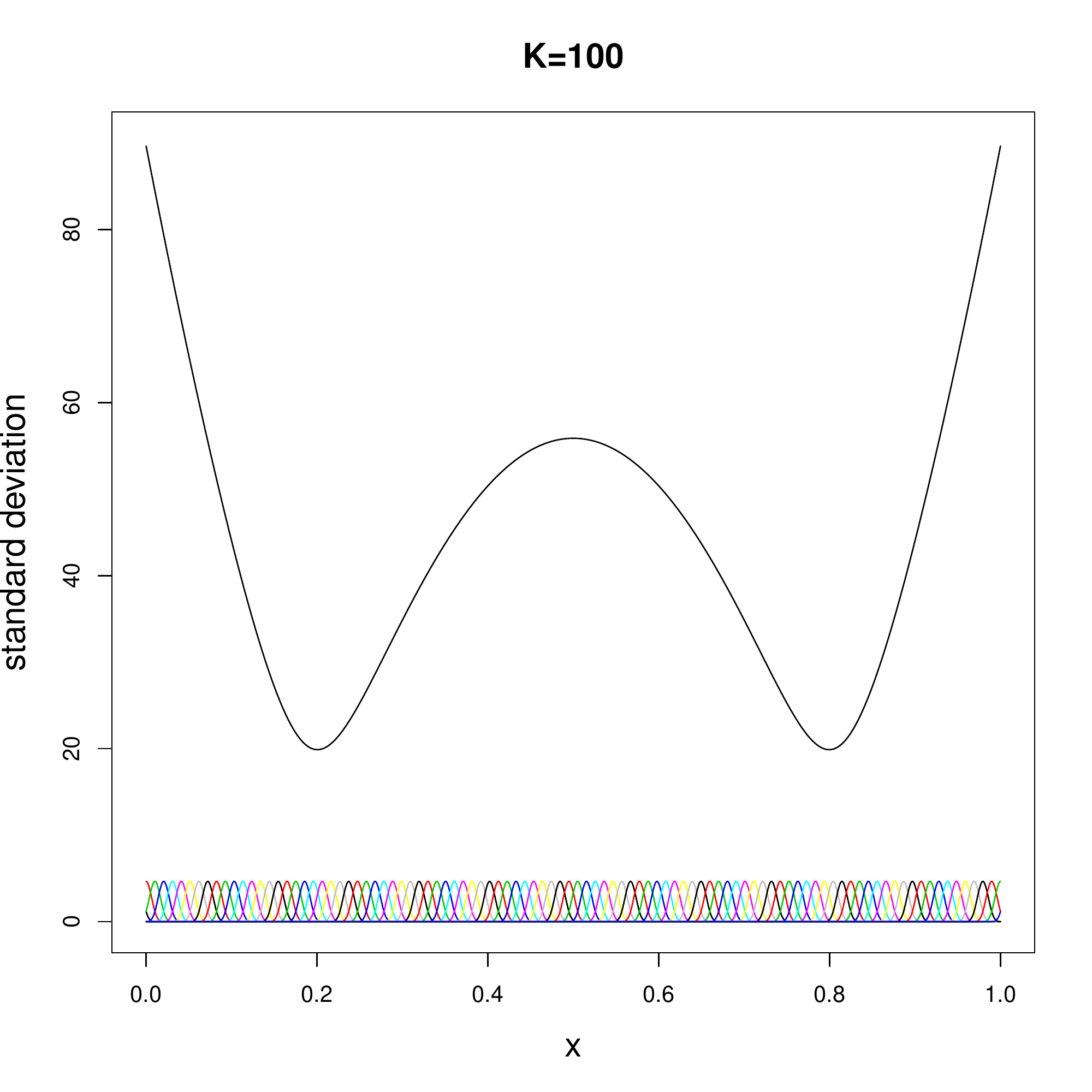}
}
\caption{The scaling issue. The two panels show the marginal standard deviation of $\bm f(\bm x)=\bm B \bm \beta$ for varying dimension of the basis $\bm B$, $K=\{50,100\}$, where $\bm B$ is a matrix of cubic B-splines (coloured lines) defined over the interval $(0,1)$ and $\bm \beta$ is an IGMRF of order 2. The standard deviation (black line) is calculated as the squared root diagonal entries of $\bm B \bm\Sigma^{*}\bm{B^{\textsf{T}}}$, with $\tau_{\beta}=1$. 
}
\label{fig:scaling-issue}
\end{figure}

The marginal variance of the IGMRF (\ref{eq:igmrf}), given by the diagonal elements of $\Sigma^{*}=\tau_{\beta}^{-1} \bm R^{-1}$, depends on $K$. We denote this as the \textit{scaling issue} \citep{sorbye-2013}, meaning that the amount of deviation from the base model depends on the number of knots. This is illustrated in Figure \ref{fig:scaling-issue}, where the two panels report the marginal standard deviation of the smooth $f(\bm x)=\bm B \bm \beta$, for $K=\{50,100\}$. On the other hand, results (not shown here) show that the degree of the B-splines has little or no impact on the marginal variance of $\bm B \bm \beta$, especially when $K$ is large enough, say $K>50$.

\subsection{Degrees of freedom}

The scaling issue can be avoided if we consider building priors on the number of effective degrees of freedom \citep{hastie-1990-book}, $
d=\text{tr} \left\{\left(\boldsymbol{B}^{\textsf{T}}\boldsymbol{B} + \frac{\tau_{\beta}}{\tau_{\epsilon}}\boldsymbol{R}\right)^{-1} \boldsymbol{B}^{\textsf{T}}\boldsymbol{B}\right\}$. 
If we think of the smooth $f(\bm x)=\bm B \bm \beta$ as a polynomial, then $d$ can be thought of as the degree of this polynomial. In presence of prior information on the degree of an equivalent polynomial, it seems a sensible approach to design a prior for $d$, $\pi(d)$, instead of $\tau_{\beta}$.

A fundamental issue when building $\pi(d)$ is that $d$ depends on both precisions $\tau_{\beta}$ and $\tau_{\epsilon}$. The former regulates the number of effective degrees of freedom, conditionally on the latter. When $\tau_{\epsilon}$ is known, the construction of $\pi(d)$ can be based on the prior $\pi(\tau_{\beta})$ (see section \ref{sec:pcprior-df-pspline}). When $\tau_{\epsilon}$ is unknown, the prior $\pi(d)$ will be specified in terms of the joint $\pi(\tau_{\beta}|\tau_{\epsilon})\pi(\tau_{\epsilon})$, following a fully Bayesian approach (see section \ref{sec:joint-pcprior}).

The degrees of freedom can be reduced to 
\begin{equation}
d = \text{tr} \left\{ \left(  \boldsymbol{I} + \frac{\tau_{\beta}}{\tau_{\epsilon}}\boldsymbol{R} (\boldsymbol{B}^{\textsf{T}}\boldsymbol{B})^{-1} \right)^{-1}\right\}  =\sum_{k=1}^{K} \frac{1}{1+\frac{\tau_{\beta}}{\tau_{\epsilon}} v_k},
\label{eq:df-sum}
\end{equation}
where $v_1,...,v_K$ are the eigenvalues of $ \boldsymbol{R}(\boldsymbol{B}^{\textsf{T}}\boldsymbol{B})^{-1}$, whose null space has dimension $r$ (the rank deficiency of $\bm{R}$). 
When the factor $\tau_{\beta}/\tau_{\epsilon}$ goes to infinity we obtain the minimum number of degrees of freedom, $d=r$. 
When $\tau_{\beta}/\tau_{\epsilon}$ goes to zero we obtain the maximum number of degrees of freedom, $K$, corresponding to the most flexible model under the assumed IGMRF.

The prior $\pi(d)$ depends on the eigenvalues of $\boldsymbol{R}(\boldsymbol{B}^{\textsf{T}}\boldsymbol{B})^{-1}$, hence on the choice of $\bm B$. Hereafter, $\bm B$ will be referred to simply as \textit{design}, because it is determined by both the assumptions made by the user (location and number of knots, order of the B-splines) and the assumptions purely made by design (location and number of observations along the covariate domain). Since the degrees of freedom depend on $\bm B$, it follows that $\pi(d)$ automatically accounts for the design. This will be discussed in detail in section \ref{sec:invariance}.

\section{PC priors for P-splines}
\label{sec:pcprior-review}

The PC prior framework by \cite{pcprior} introduces a new concept for building priors in hierarchical additive models, where the latent structure is given by the sum of a number of model components described by a small number of flexibility parameters. Each model component is seen as a flexible extension of a base model. For instance, $\tau_{\beta}$ is a flexibility parameter for the IGMRF component $\pi(\bm \beta|\tau_{\beta})$ and a natural base model corresponds to $\pi(\bm \beta|\tau_{\beta}^{-1}=0)$. Below, the four principles underpinning the construction of a PC prior for $\tau_{\beta}$ are reviewed, following \cite{pcprior}.

\begin{itemize}

\item \textit{Parsimony}. A simple model should be preferred unless there is enough evidence for a more flexible one. Under this principle, the prior probability mass assigned to models of increasing complexity should decay as their distance from the base model (measured in terms of model complexity) increases. In Bayesian P-splines, the IGMRF prior operates on $\bm\beta$ (the object of inference) but we can extend the notion of base and flexible model to the \textit{spline-modelled} function; we denote with $\bm f_0=\bm B \bm \beta_0$ the base model, which is a polynomial of degree $r-1$, and with $\bm f=\bm B \bm \beta$ the flexible model, which reflects any deviation from such polynomial. 

\item \textit{Measure of complexity}. The Kullback-Leibler divergence \citep[KLD, ][]{kld-1951} is assumed to evaluate the distance, $\delta$, between the complexities of two different models. We use $\text{KLD}(\bm f || \bm f_0)$ to denote the increased complexity of the flexible model $\bm f$ with respect to the base model $\bm f_0$. Since $\bm B$ is fixed by design, it is enough to evaluate $\text{KLD}(\bm \beta|| \bm \beta_0)$. 
Let $\tau_{\beta_0}$ and $\tau_{\beta}$ be the precisions of the base and flexible model, respectively, it can be shown that $\text{KLD}(\bm \beta|| \bm \beta_0)$ goes to $\frac{\tau_{\beta_0}K}{2 \tau_{\beta}}$, for $\tau_{\beta}$ much lower than $\tau_{\beta_0}$ and $\tau_{\beta_0} \to \infty$; see a proof in \cite{pcprior}. Finally, for convenience we take the transformation $\delta=\sqrt{2 \text{KLD}(\bm \beta||\bm \beta_0)}= \sqrt{{\tau_{\beta_0}K}/{ \tau_{\beta}}}$.  

\item \textit{Constant rate penalization}. Flexible models are penalized by a constant decay rate for increasing $\delta$. Following this principle, the PC prior is defined as an exponential distribution on the distance scale, $\pi_{PC}(\delta)=\lambda \exp(-\lambda \delta)$, with constant rate $\lambda$. It follows that the mode of a PC prior is always at the base model. By a change of variable and setting the rate $\lambda=\theta/\sqrt{K \tau_{\beta_0}}$, \cite{pcprior} obtain the PC prior for $\tau_{\beta}$ as,
\begin{eqnarray}
\pi_{PC}(\tau_{\beta}) & = & \lambda \exp\left(-\lambda \sqrt{{ \tau_{\beta_0}K}/{ \tau_{\beta}}} \right) \left| \frac{\partial \sqrt{{\tau_{\beta_0}K}/{ \tau_{\beta}}}}{\partial \tau_{\beta}}\right| \nonumber \\
& = & \frac{\theta}{2} \tau_{\beta}^{-{3}/{2}} \exp\left( -{\theta}/{\sqrt{\tau_{\beta}}}\right),
\label{gumbel}
\end{eqnarray}
which is a $\text{Gumbel}(1/2, \theta)$ type 2 distribution, $\theta>0$.

\item \textit{User-defined scaling}. Often, the user has an idea about the size of an interpretable transformation of the original parameter $\tau_{\beta}$, say $h(\tau_{\beta})$ (e.g. degrees of freedom).
In this case the user may elicit an upper bound $U$ for $h(\tau_{\beta})$ and set a prior probability $\alpha$ for the tail event, i.e. $\alpha=Pr(h(\tau_{\beta})>U)$. \cite{pcprior} suggest to bound the marginal standard deviation, $1/\sqrt{\tau_{\beta}}$. To obtain $\theta$ in (\ref{gumbel}) it is enough to specify $(U,\alpha)$ and solve $Pr(1/\sqrt{\tau_{\beta}} > U) = \alpha$ for $\theta$, which yields $\theta=-\log(\alpha)/U$.
\end{itemize}

PC priors can be helpful as \textit{default} priors in complex hierarchical models where, typically, ``it is difficult to elicit information about structural parameters that are further down the model hierarchy'' \citep{pcprior}. In addition, the user-defined scaling approach enables to build \textit{informative} priors for the original parameter $\tau_{\beta}$ or for a property of the associated model component, by tuning two intuitive parameters $U$ and $\alpha$. 
In the next section we introduce a new scaling approach to derive the PC prior for the degrees of freedom of a P-spline model component $\bm B \bm \beta$. Other approaches might be possible though. For instance, recently \cite{klein-2015} proposed PC priors for the scale (or range of variation) of $\bm B \bm \beta$, and showed via simulation that these outperformed the Gamma family in cases where the data are weakly informative and/or the size of the effects is close to the base model.

\section{PC-priors for degrees of freedom} 
\label{sec:pcprior-df-pspline}

\subsection{A new scaling approach}
\label{sec:scaling}

With no loss of generality, we derive the PC prior for degrees of freedom and study its properties under the assumption that the noise precision $\tau_{\epsilon}$ is known. Given $\tau_{\epsilon}$, denote as $d(\tau_{\beta})=\sum_{k=1}^{K} \frac{1}{1+\frac{\tau_{\beta}}{\tau_{\epsilon}} v_k}$ the function mapping the precision $\tau_{\beta}$ into the number of effective degrees of freedom, following (\ref{eq:df-sum}). (Hereafter $d(\tau_{\beta})$ will be referred to as the mapping). Figure \ref{fig:mapping} shows the mapping $d$ in the log precision scale for $\tau_{\epsilon}=1$ and various designs (left panel) and for a specific design and varying $\tau_{\epsilon}$ (right panel). 

We introduce a new user-defined scaling operating not directly on $\tau_{\beta}$, but on $d(\tau_{\beta})$. 
Let $U$ be an upper bound for $d(\tau_{\beta})$ and $\alpha$ a (small) probability associated to the tail event,
\begin{equation*}
\alpha = Pr(d(\tau_{\beta})>U) = Pr(\tau_{\beta} < d^{-1}(U)) =  F(d^{-1}(U))
\label{condition} 
\end{equation*} 
where $F$ is the c.d.f of the $\text{Gumbel}(1/2, \theta)$ type 2 distribution.
The PC prior resulting from this new scaling is a Gumbel type 2 as in (\ref{gumbel}) with $\theta=-\log(\alpha)\sqrt{d^{-1}(U)}$. 
In the following, $\pi_{PC}(d)$ will denote the \textit{induced PC prior for degrees of freedom}, with $U \in (r,K)$ and $\alpha \in (0,1)$ the parameters specifying the distribution. 


\subsection{Invariance under design}
\label{sec:invariance}

\begin{figure}
\centerline{\includegraphics[width=.5\textwidth]{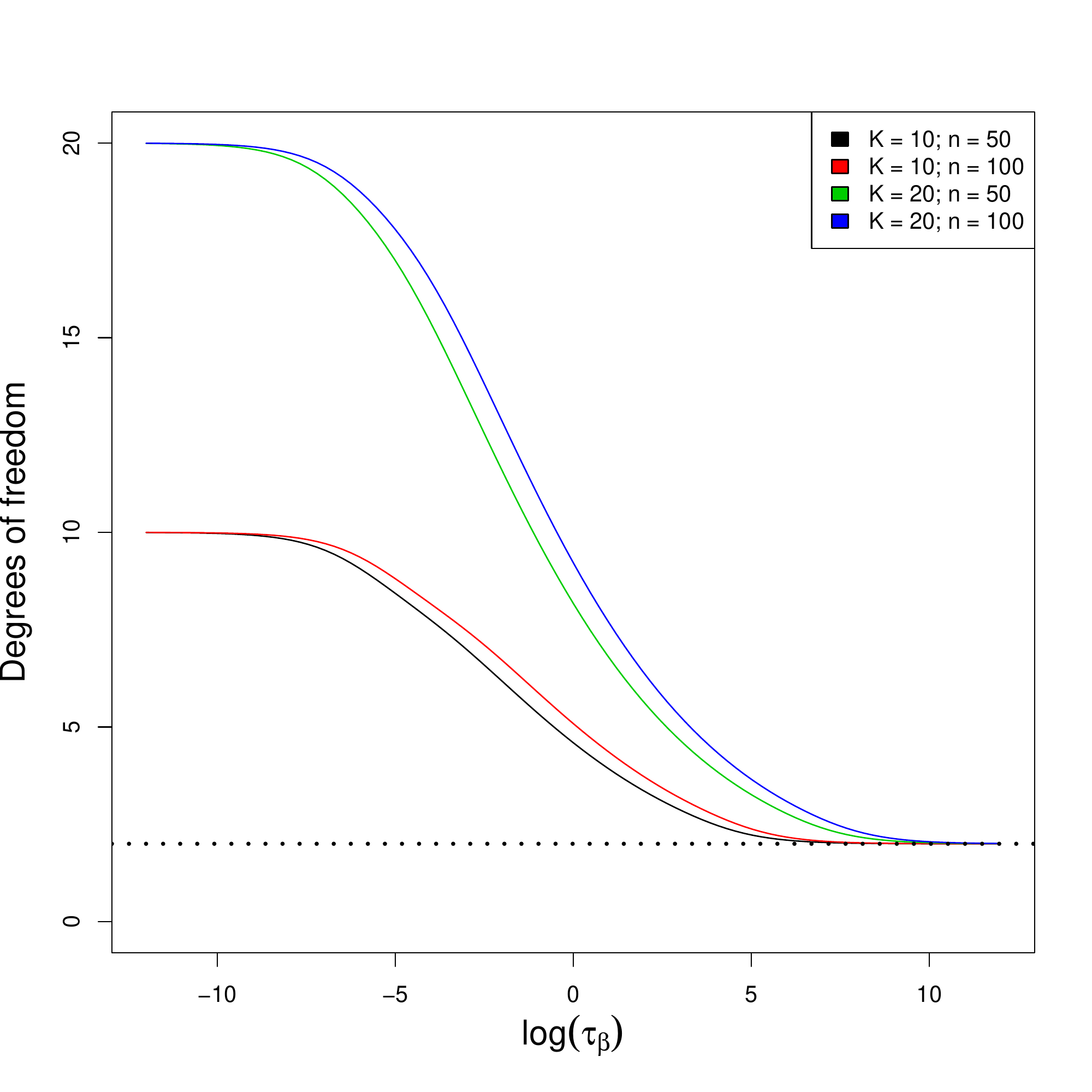}
\includegraphics[width=.5\textwidth]{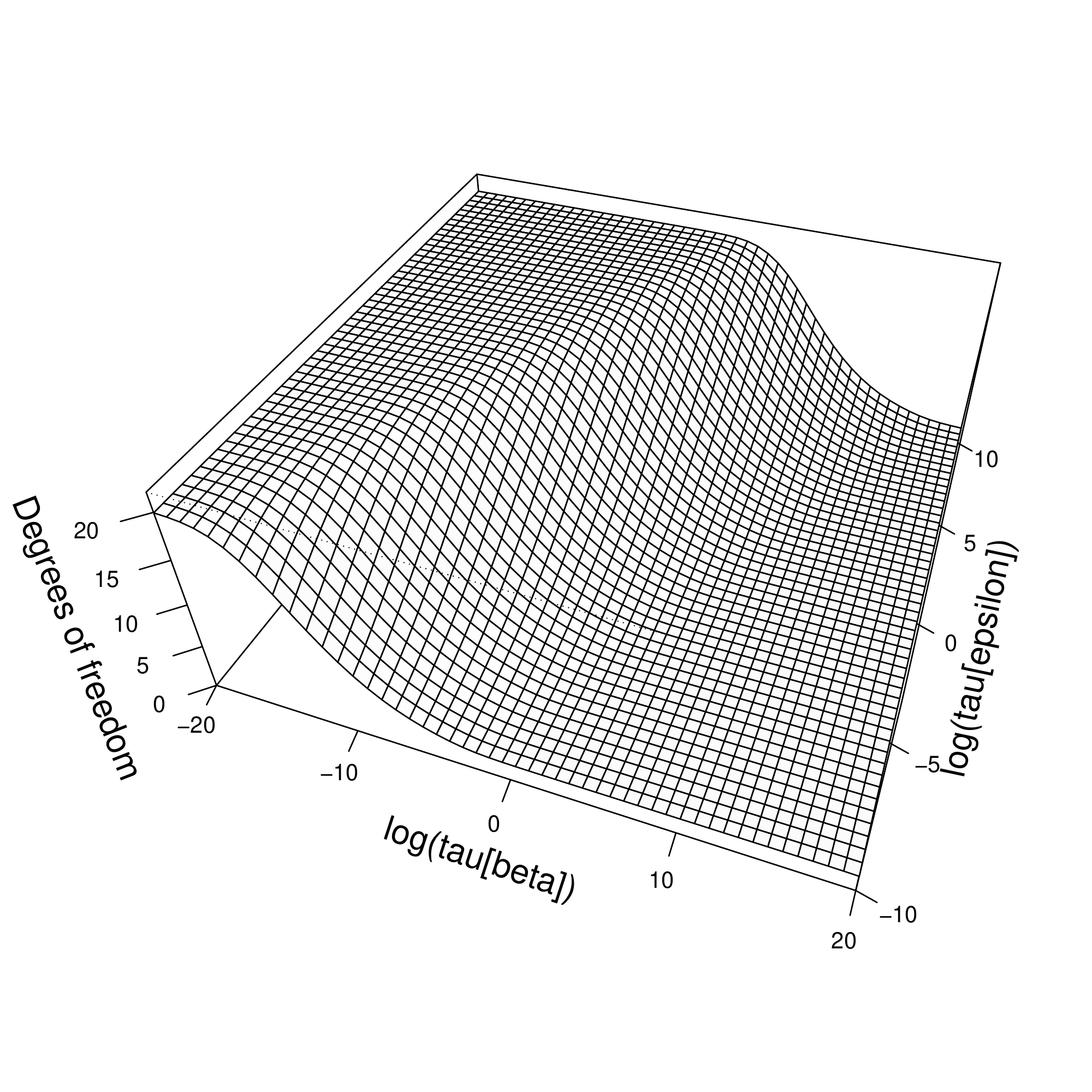}
}
\caption{Mapping the degrees of freedom. The plot on the left shows the mapping $d$ in the $\log(\tau_{\beta})$ scale, conditional on $\tau_{\epsilon}=1$, for four designs (choices of $K$ and $n$). 
The dotted horizontal line at $d=2$ indicates the base model (assuming an IGMRF of order 2 on the spline coefficients). The plot on the right shows $d$ as a function of both $\tau_{\epsilon}$ and $\tau_{\beta}$, for the specific design $\{K=20,n=50\}$. 
}
\label{fig:mapping}
\end{figure}

\begin{figure}
\centerline{
\includegraphics[width=.5\textwidth]{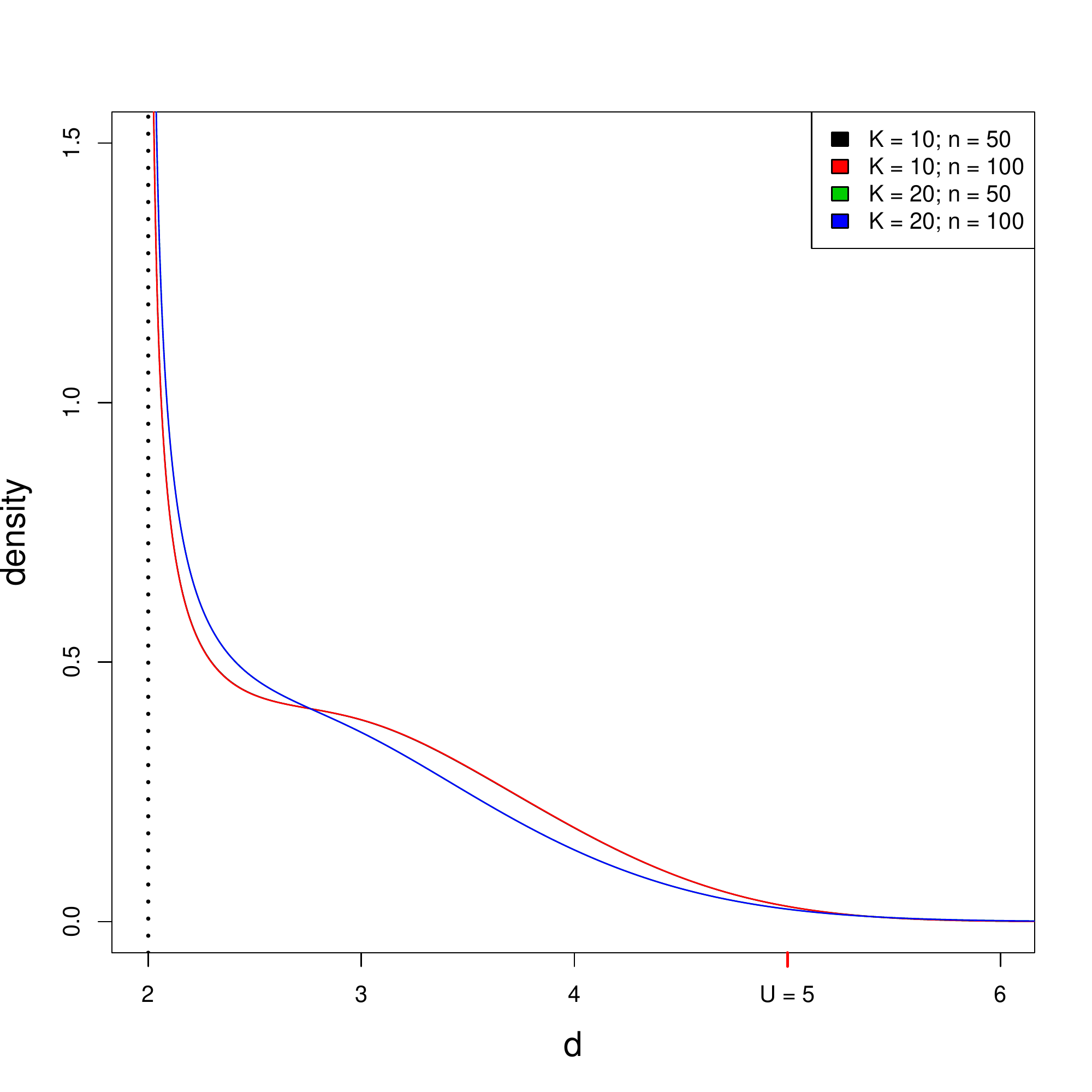}
\includegraphics[width=.5\textwidth]{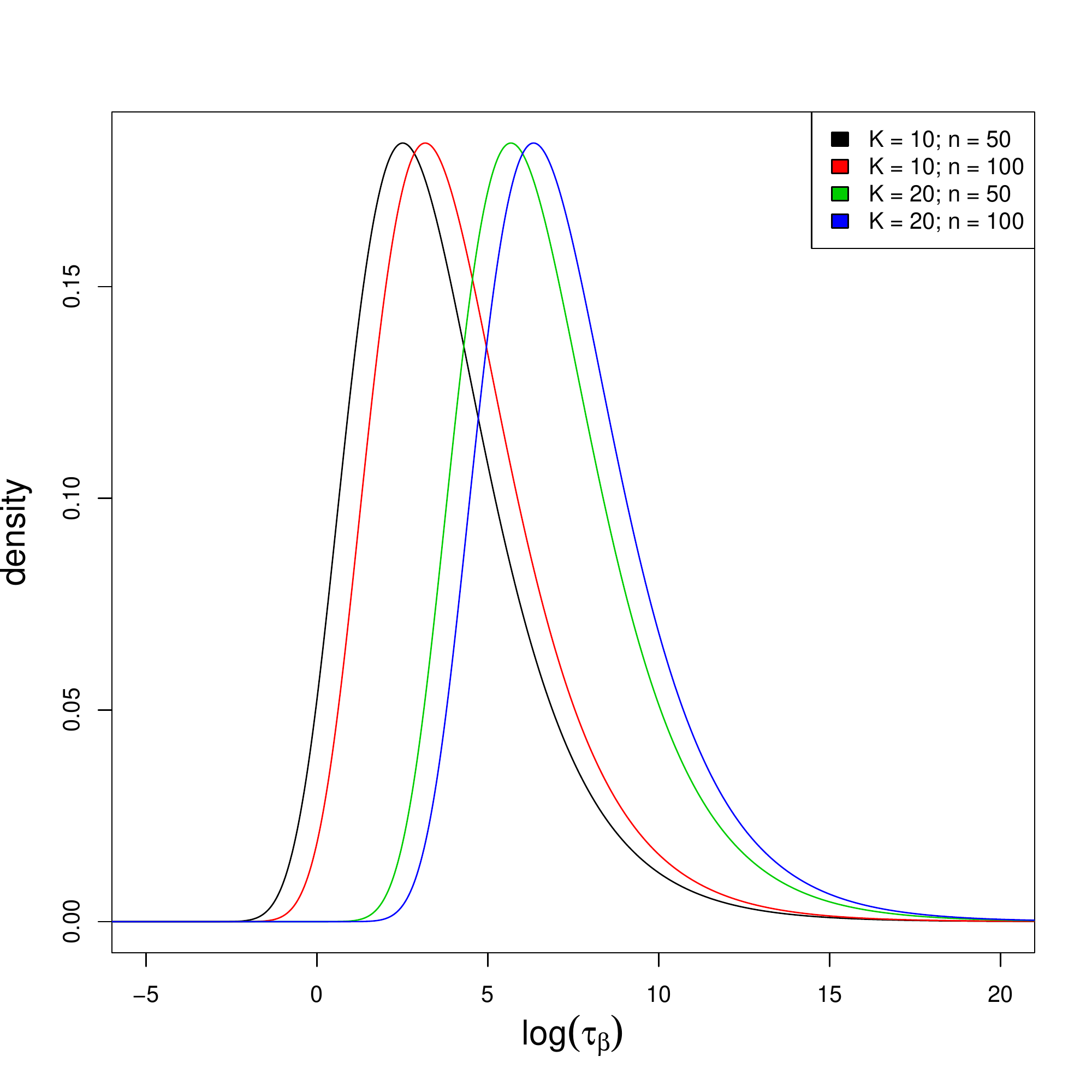}
}
\caption{Invariance under design. Both panels show the PC prior density $\pi_{PC}(d)$, for $(U=5,\alpha=0.01)$ and $ \tau_{\epsilon}=1$, in two different scales: $d$ (left panel) and $\log(\tau_{\beta})$ (right panel). In the left panel, the dotted vertical line at $d=2$ indicates the base model (assuming an IGMRF of order 2 on $\bm \beta$), while the red tick indicates the upper bound for degrees of freedom. Even if the PC prior density changes for varying $K$ and $n$, the probability assigned to $d>5$ is always $0.01$. In this particular example, the density change between $n=50$ and $n=100$ is evident in the $\log(\tau_{\beta})$ scale (right panel), but not in the $d$ scale (left panel), where the black and red lines, as well as the green and blue lines, appear superimposed (however, they are not the same because the eigenvalues of $ \boldsymbol{R}(\boldsymbol{B}^{\textsf{T}}\boldsymbol{B})^{-1}$ are different). }
\label{fig:pcprior}
\end{figure}

While the PC prior for $\tau_{\beta}$ in (\ref{gumbel}) does not take into account any information regarding the adopted design, the induced PC prior for $d$ does. Indeed, different designs return different mappings $d$ (see the left panel in Figure \ref{fig:mapping}), which implies a desirable property: the PC prior $\pi_{PC}(d)$ is \textit{invariant under design}; the term invariant here applies to the interpretation of the PC prior in terms of degrees of freedom, not to the density. 
Figure \ref{fig:pcprior} illustrates this property. The density of $\pi_{PC}(d)$, with parameters $(U=5,\alpha=0.01)$, is displayed both in the $d$ scale (left panel) and $log(\tau_{\beta})$ scale (right panel), for different designs. By changing the design, the range of $d$ also changes and the density $\pi_{PC}(d)$ adapts accordingly (Figure \ref{fig:pcprior} left). However, even if $\pi_{PC}(d)$ materializes differently in different designs, the probability mass assigned to $d>U$ is always $\alpha$. In the $\log(\tau_{\beta})$ scale, the location of the PC prior is shifted between different designs (Figure \ref{fig:pcprior} right). In our opinion, this shows well how difficult it is to define priors for degrees of freedom in the original scale: in this case, the user would have to figure out the correct location of $\pi_{PC}(\log(\tau_{\beta}))$ and shift it according to the adopted design.

The PC prior $\pi_{PC}(d)$ plotted in the left panel of Figure \ref{fig:pcprior} has been obtained numerically. Let $\pi_{PC}(\tau_{\beta}')$ be the PC prior (\ref{gumbel}) evaluated at some predefined $\tau_{\beta}'>0$ (a convenient approach is to take $\log(\tau_{\beta}')$ on a regular grid inside $(-20,20)$) and $d'=d(\tau_{\beta}')$ be the associated degrees of freedom computed by (\ref{eq:df-sum}). The induced PC prior evaluated at $d'$ is $\pi_{PC}(d^{-1}(d'))\left| \frac{\partial d^{-1}(d')}{\partial d'} \right|$.

\subsection{Behaviour near the base model}
According to \cite{pcprior}, the prior $\pi(\delta)$, where $\delta$ is the distance from the base model, is said to overfit, or to force overfitting, if $\pi(0)=0$. Theorem 1 in \cite{pcprior} states that, if $\pi(\tau_{\beta})$ is an absolutely continuous prior for the IGMRF precision $\tau_{\beta}$, with $E(\tau_{\beta}) < \infty$, this prior overfits. The commonly used $\text{Gamma}(a,b)$, $a,b>0$, with $a/b< \infty$ falls in this class of overfitting priors.

PC priors avoid over-fitting by construction: by applying the first three principles outlined in section \ref{sec:pcprior-review} the mode of a PC prior is always at the base model. The fourth principle essentially allows the user to specify the penalty for increasing distances from the base model. 
The following result describes the behaviour near the base model for both the PC prior for degrees of freedom and the distribution of the degrees of freedom induced by a $\text{Gamma}(a,b)$ on $\tau_{\beta}$, which we denote as $\pi_{G}(d)$. (In the result below, we consider $\theta=-\log(\alpha)\sqrt{d^{-1}(U)}$, where $d$ is the mapping given a positive and finite $\tau_{\epsilon}$.)

\textbf{Result}. 
Let $r$ be the dimension of the null space of $\bm R$ in (\ref{eq:igmrf}), then $\pi_{PC}(d) \to \infty$ as $d \to r$, for $\theta>0$, and $\pi_{G}(d) \to 0$ as $d \to r$, for $a,b>0$.  

\begin{figure}
\centerline{
\includegraphics[width=.5\textwidth]{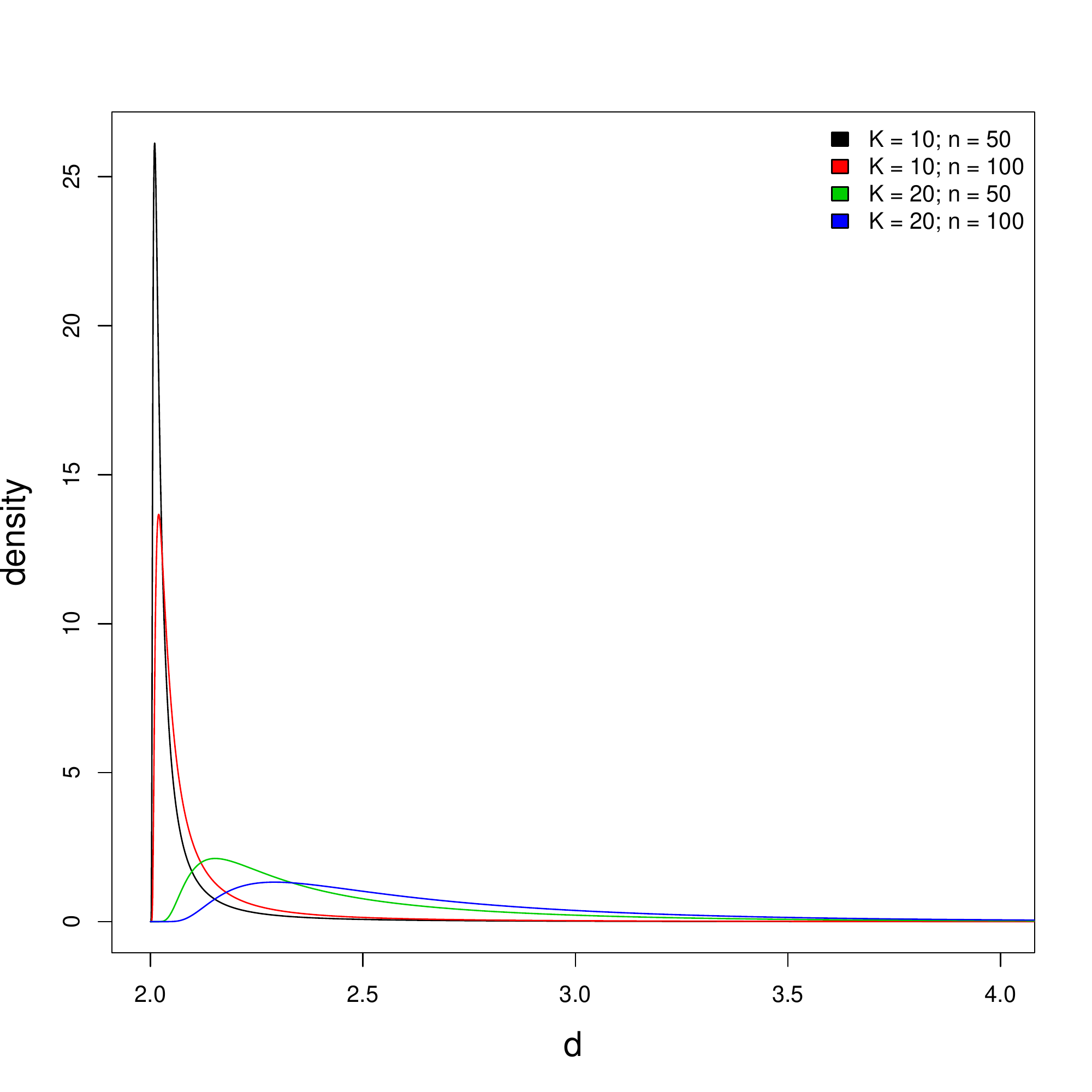}
\includegraphics[width=.5\textwidth]{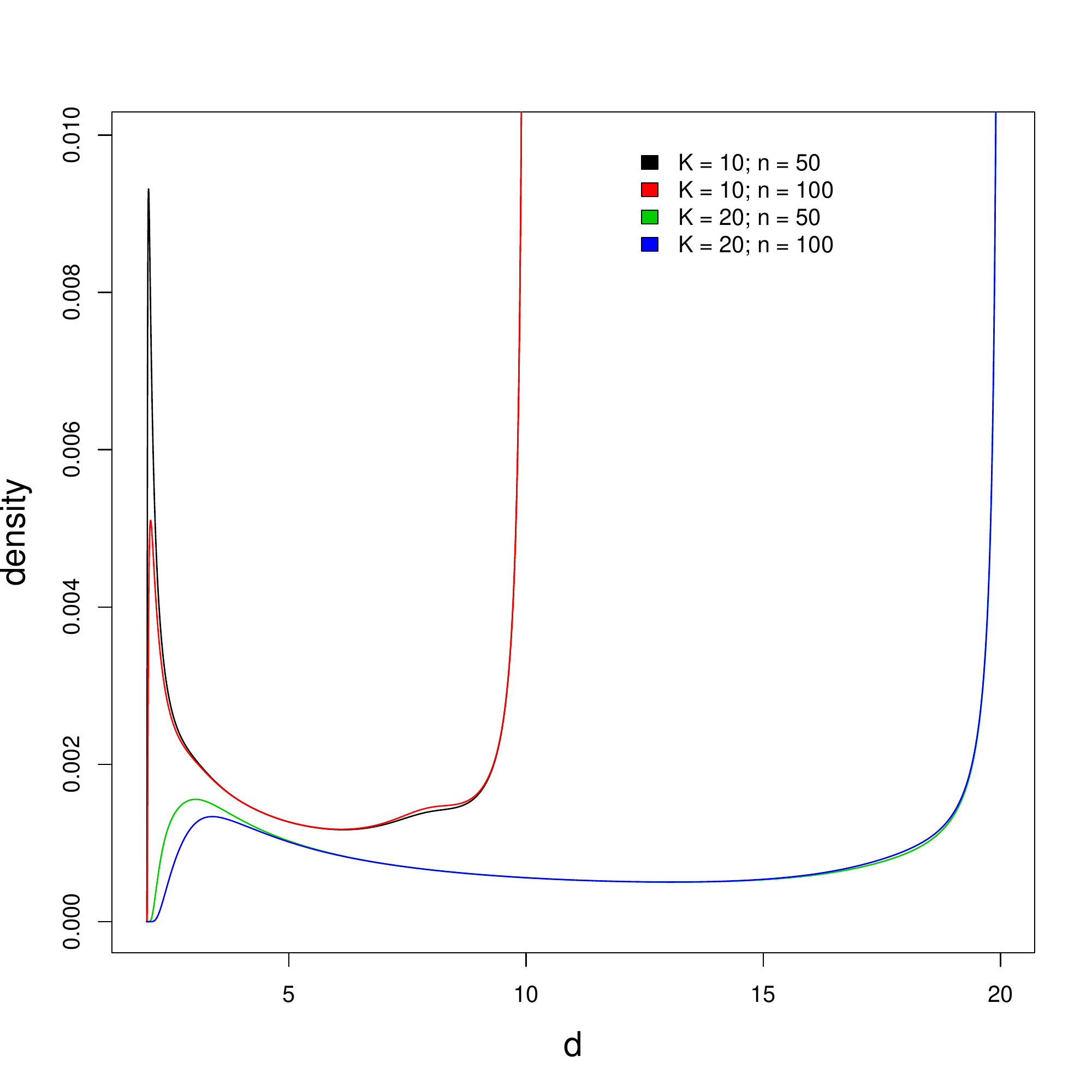}}
\caption{The distribution of the number of effective degrees of freedom induced by a Gamma prior with $a=1,b=5e-4$ (left panel) and $a=10^{-3},b= 10^{-3}$ (right panel), under four different designs. The base model is at $d=2$, assuming an IGMRF prior of order 2 on $\bm \beta$.}
\label{fig:edf-gamma}
\end{figure}

The proof is given in appendix \ref{app:A1}. The density $\pi_{PC}(d)$ goes to infinity as approaching the base model, avoiding over-fitting. Instead, the Gamma-induced $\pi_{G}(d)$ does not prevent over-fitting as it repulses the base model. In Figure \ref{fig:edf-gamma}, the Gamma-induced priors with $a=1,b=5 \cdot 10^{-4}$ (left panel) and $a=10^{-3}, b=10^{-3}$ (right panel) are displayed under four different designs. These two different priors have different interpretations in terms of degrees of freedom: the first favours over-smoothing while the latter favours over-fitting. For both choices of $a$ and $b$, the base model is repulsed at a different rate according to design. Indeed, for $a$ and $b$ fixed, the density $\pi_G(d)$ clearly changes with design: in general, Gamma priors are not invariant under design as a consequence of the scaling issue discussed in section \ref{sec:pspline}.

\section{P-splines with a joint prior on $(\tau_{\beta}, \tau_{\epsilon})$}
\label{sec:joint-pcprior} 

So far we have worked under the assumption of known noise precision $\tau_{\epsilon}$. 
The number of effective degrees of freedom is a function of $\tau_{\beta}$, which scales differently according to the level of noise present in the data (see the right panel in Figure \ref{fig:mapping}). 
Knowing $\tau_{\epsilon}$ is then crucial to scale the PC prior correctly, in order to guarantee the upper bound for degrees of freedom specified by the user. 

The noise precision is typically unknown in applications. One could estimate $\tau_{\epsilon}$ from the data and then specify the PC prior for $d$ conditional on this estimate; this strategy has been proposed by \cite{fong-2010} to define Gamma-induced priors for degrees of freedom. We, instead, adopt a fully Bayesian model,
\begin{eqnarray}
\boldsymbol{y}| \boldsymbol{\beta}, \tau_{\epsilon} & \sim & N(\boldsymbol{B} \boldsymbol{\beta}, \tau_{\epsilon}^{-1} \boldsymbol{I}) \label{eq:mod-lik} \\
\boldsymbol{\beta}| \tau_{\beta} & \sim & N(0, \tau_{\beta}^{-1} \boldsymbol{R}^{-1}) \label{eq:mod-beta} \\
\tau_{\beta} | \tau_{\epsilon} & \sim & \text{Gumbel}(1/2, \theta(\tau_{\epsilon})) \label{eq:mod-cond-prior}\\
\tau_{\epsilon} & \sim & \pi(\tau_{\epsilon}) \propto 1/\tau_{\epsilon},
\label{eq:mod-prior-taue}
\end{eqnarray}
where (\ref{eq:mod-cond-prior}) and (\ref{eq:mod-prior-taue}) specify a joint prior $\pi(\tau_{\beta}, \tau_{\epsilon})=\pi_{PC}(\tau_{\beta}|\tau_{\epsilon})\pi(\tau_{\epsilon})$. The scaling parameter in (\ref{eq:mod-cond-prior}) is given by $\theta(\tau_{\epsilon})=-\log(\alpha)\sqrt{d^{-1}(U|\tau_{\epsilon})}$, where $d(\cdot|\tau_{\epsilon})$ is the mapping conditional on the random noise precision $\tau_{\epsilon}$. 
We use the improper $\pi(\tau_{\epsilon}) \propto 1/\tau_{\epsilon}$ since the data usually contain sufficient information with respect to $\tau_{\epsilon}$. The joint prior in (\ref{eq:mod-cond-prior}) and (\ref{eq:mod-prior-taue}) corresponds to the induced PC prior for $d$ conditional on a random $\tau_{\epsilon}$, with $U \in (r,K)$ and $\alpha \in (0,1)$ the parameters specifying the distribution.

We developed a Markov chain Monte Carlo (MCMC) algorithm to fit model (\ref{eq:mod-lik}) to (\ref{eq:mod-prior-taue}); see pseudo-code reported in Appendix \ref{app:A2} algorithm \ref{mcmc1}. The algorithm includes a Metropolis-Hasting step to jointly update $(\tau_{\epsilon}, \tau_{\beta}, \bm{\beta})$. In our experience, block updating ensures good mixing and fast convergence of the proposed MCMC algorithms; see \cite{rue-2005} for details on block updating in hierarchical models with GMRF components. A brief description on how algorithm \ref{mcmc1} works follows. At iteration $j$, both precision parameters (here denoted simply as $\tau$) are sampled from the proposal distribution adopted in \cite{held-rue-2002}: $q(\tau^{*}|\tau^{(j-1)})=  t \tau^{(j-1)}$, where $\tau^{*}$ and $\tau^{(j-1)}$ are, respectively, the proposed and current values at iteration $j$, $t$ is random with density $\pi(t) \propto 1+1/t$, for $t \in \left[ 1/T, T\right]$ and $T>1$ is a tuning parameter; in our experience, setting $T$ approximately equal to $1.5$ works well in most applications. The proposed $\bm{\beta}^{*}$ is sampled from the full conditional of the spline coefficients given the proposed $\tau_{\epsilon}^{*}$ and $\tau_{\beta}^{*}$. To compute the acceptance probability for $(\tau_{\epsilon}^{*}, \tau_{\beta}^{*}, \bm{\beta}^{*})$ in step (\ref{algo1:accept}) of algorithm \ref{mcmc1}, the joint posterior 
\begin{equation}
\pi(\tau_{\epsilon}^{*}, \tau_{\beta}^{*}| \bm{y})  \propto  \frac{\pi(\bm{y}|\bm{\beta}^{*}, \tau_{\epsilon}^{*}) \pi(\bm{\beta}^{*}|\tau_{\beta}^{*}) \pi_{PC}(\tau_{\beta}^{*}|\tau_{\epsilon}^{*}) \pi(\tau_{\epsilon}^{*})}{
\pi(\bm \beta^{*}|\tau_{\beta}^{*},\bm y)}
\label{eq:accept-numerator}
\end{equation}
needs to be evaluated. It is crucial to rescale $\pi_{PC}(\tau_{\beta}|\tau_{\epsilon})$ according to the proposed $\tau_{\epsilon}^{*}$ before calculating (\ref{eq:accept-numerator}). This implies re-evaluating the inverse mapping $d^{-1}(\cdot|\tau_{\epsilon})$, given $\tau_{\epsilon}^{*}$, and recomputing $\theta(\tau_{\epsilon}^{*})$ at each iteration.

\subsection{Additive P-splines}
\label{sec:joint-pcprior-identifiability}
We now focus on an additive P-spline modelling framework, where the linear predictor is the sum of a number of smooth functions. Let $\bm y$ be a Gaussian response and $\bm x_j$, $j=1,...,J$, be a set of $J$ continuous covariates, the model is 
\begin{eqnarray}
\bm y & = & \sum_{j=1}^{J} f_j(\bm x_j) + \bm \epsilon \ \ \ \ ; \ \ \ \ \bm \epsilon \sim N(0,\tau_{\epsilon}^{-1}) \nonumber \\
f_j(\bm x_j) & = & \bm B_j \bm \beta_j, \nonumber \\
\boldsymbol{\beta_j}| \tau_{\beta_j} & \sim & N(0, \tau_{\beta_j}^{-1} \boldsymbol{R}^{-1}) \nonumber
\end{eqnarray}
where $\bm B_j$ is the $n \text{ x } K_j$ B-spline basis matrix and $\bm \beta_j$ the vector of spline coefficients associated to the smooth function $f_j$; with no loss of generality, we consider the same number of knots $\forall j$, yielding $K=K_j$, $j=1,...,J$. We assume the joint prior $\prod_{j=1}^{J}\pi_{PC}(\tau_{\beta_j}|\tau_{\epsilon})\pi(\tau_{\epsilon})$, where $\pi(\tau_{\epsilon}) \propto 1/\tau_{\epsilon}$ and $\pi_{PC}(\tau_{\beta_j}|\tau_{\epsilon})= \text{Gumbel}(1/2, \theta_j(\tau_{\epsilon}))$, $j=1,...,J$. The scaling parameter $\theta_j(\tau_{\epsilon})=-\log(\alpha_j)\sqrt{d_j^{-1}(U_j|\tau_{\epsilon})}$, where $(U_j,\alpha_j)$ are the parameters calibrating the induced PC prior for the degrees of freedom of $f_j$. The mapping for the degrees of freedom of $f_j$ is given by $d_j(\tau_{\beta_j}|\tau_{\epsilon})=\text{tr}\left\{\left(\boldsymbol{B_j}^{\textsf{T}}\boldsymbol{B_j} + \frac{\tau_{\beta_j}}{\tau_{\epsilon}}\boldsymbol{R}\right)^{-1} \boldsymbol{B_j}^{\textsf{T}}\boldsymbol{B_j}\right\}$.  

Identifiability constraints are important in additive P-splines. The IGMRF prior on $\bm \beta_j$ controls deviations of the smooth term $\bm B_j \bm \beta_j$ from a polynomial base model; in the following, for simplicity, we consider an IGMRF of order 2 (which forces shrinkage towards a linear base model). All smooths $\bm B_j \bm \beta_j$ include the linear base model, thus they all compete to capture the mean of the data. To ensure identifiability we adopt the following re-parametrization, 
\begin{equation}
\bm y = \mu + \sum_{j=1}^{J} \bm x_j \gamma_j + \sum_{j=1}^{J}\bm B_j \bm \beta^{ULC}_j + \bm \epsilon, 
\label{eq:mod-add-ULC}
\end{equation}
where $\mu$ is the intercept, $\gamma_j$ is the slope coefficient for covariate $\bm x_j$ and $\bm \beta^{ULC}_j$ are the spline coefficients $\bm \beta_j$ under the two following linear constraints: $\bm \left[\bm c^{\textsf{T}} \bm B_j \right] \bm \beta_j=0$ and $\left[\bm l^{\textsf{T}} \bm B_j \right] \bm \beta_j=0$, with \textit{constant} vector $\bm c=\bm 1_n$ and \textit{line} vector $\bm l=\left[1, 2,...,n\right]^{\textsf{T}}$. In this way, the smooth term $\bm B_j \bm \beta^{ULC}_j$ captures residual variations from the linear base model $\mu + \gamma_j \bm x_j$. In other words, the constrained model (\ref{eq:mod-add-ULC}) allows each smooth component to be identified, by separating the linear and flexible terms which coexist in each $\bm B_j \bm \beta_j$. Model (\ref{eq:mod-add-ULC}) can be expressed in compact form as 
\begin{equation}
\bm y = \bm X \bm \gamma + \bm B \bm \beta^{ULC} + \bm\epsilon,
\label{eq:mod-add-ULC-compact}
\end{equation} 
where $\bm B=\left[ \bm B_1:...: \bm B_j\right]$ is the $n \text{ x } (KJ)$ joint basis matrix, $\bm \beta^{ULC}$ is the joint vector of spline coefficients subject to the linear constraints, $\bm X=\left[\bm 1_n: \bm x_1:... :\bm x_J \right]$ is the $n \text{ x } (J+1)$ matrix of covariates with an additional column of ones for the intercept term and $\bm \gamma$ is the vector of fixed effects. We assume $\bm \gamma \sim N(\bm 0, \tau_{\gamma} \bm I_{J+1})$ with a small precision, e.g. $\tau_{\gamma}=10^{-4}$, as a prior for the fixed effects. Other covariates can be added to $\bm X$ in model (\ref{eq:mod-add-ULC-compact}), if we assume them to have a simple linear effect.

We wrote a block updating MCMC algorithm implementing the joint prior in the model described above; pseudo-code is given in Appendix \ref{app:A2} algorithm \ref{mcmc2}. This includes two Metropolis-Hasting steps to jointly update the blocks $(\tau_{\epsilon}, \bm \gamma)$ and $(\tau_{\beta_j}, \bm \beta_j^{ULC})$, for each $j=1,...,J$; in our experience, this scheme gives good mixing (and convergence) properties. Algorithm \ref{mcmc2} presents two main changes with respect to algorithm \ref{mcmc1}. First, rescaling the conditional PC prior $\pi_{PC}(\tau_{\beta_j}| \tau_{\epsilon})$ is no longer necessary at each iteration; $\theta(\tau_\epsilon^{*})$ must be recomputed only when $(\tau_{\epsilon}^{*}, \bm \gamma^{*})$ is accepted. Second, the spline coefficients are sampled under linear constraints. To do this we use the algorithm proposed in \cite{rue-2005} ch. 2, which samples first the unconstrained coefficients and then \textit{corrects} them for the constraints. To compute the acceptance probability for the candidate $(\tau_{\beta_j}^*, \bm \beta_j^{*ULC})$ in step (\ref{algo2:accept-beta})  of algorithm \ref{mcmc2} Appendix \ref{app:A2}, the full conditional density $\pi(\bm \beta^{*}|\tau_{\beta}^*,\bm y)$ is evaluated at the constrained $\bm \beta_j^{*ULC}$; for computational details see \cite{rue-2005}, formula (2.31).

\section{Simulation study}
\label{sec:simulation-study}

The scaling parameter $\theta$ of the PC prior (\ref{gumbel}) can be tuned in several ways. Our proposal is to select $\theta$ through assumptions on the number of effective degrees of freedom, $d=h(\tau_{\beta})$, of the function $f(\bm x)=\bm B \bm \beta$. 
This approach seems intuitive in the Gaussian case where quantity $d$ relates immediately to the degree of an equivalent polynomial (which an expert user might have prior information about). Moreover, literature on smoothing often refers to degrees of freedom as a way to summarize model complexity. 
In a recent paper, \cite{klein-2015} specify $\theta$ through assumptions on the \textit{scale}, or range of variation, of $f(\bm x)=\bm B \bm \beta$ and denote this PC prior as \textit{scale-dependent prior} (SD prior). Precisely, $\theta$ is derived numerically by requiring that $Pr(|f(x)| \leq c) \geq 1-\alpha$ for each $x$ in the covariate domain, where $\alpha \in (0,1)$ and $c$ indicates an upper bound for the scale of $f$. Both scaling approaches lead to a PC prior which is invariant under design, in the sense that the computation of $\theta$ accounts for the adopted B-spline design $\bm B$. However, the two PC priors differ regarding conditioning on the noise variance. Our PC prior, $\pi_{PC}(d)$, is a joint prior $\pi(\tau_{\beta}|\tau_{\epsilon})=\pi_{PC}(\tau_{\beta}| \tau_{\epsilon})\pi(\tau_{\epsilon})$, while the SD prior is defined unconditionally on $\tau_{\epsilon}$ as the scale of $f$ does not depend on $\tau_{\epsilon}$.

In this section we present a simulation study which investigates further the relevance of degrees of freedom for designing priors for Gaussian P-splines. The objective of our study is to assess the behaviour of our joint prior in scenarios with different noise levels, and to compare this with two alternative priors which are defined unconditionally on $\tau_{\epsilon}$, namely the conjugate Gamma prior and the SD prior. Therefore, our simulation study does not aim to generically assess the behaviour of PC priors compared to standard priors. For an extended simulation study evaluating the performance of PC priors (in particular, SD priors) compared to several alternative hyper-priors for variance parameters, in both Gaussian and non Gaussian contexts, the reader is referred to \cite{klein-2015}. Furthermore, our simulation is restricted to the Gaussian case.

We consider two different models regarding the shape of the true $f(x)$:
\begin{itemize}
\item $f_1(x)=\sin(x) ; \ \ \ \ \ \  x \in (-1,1)  $\\
\item $f_2(x)=\cos(x) ; \ \ \ \ \ \; x \in (0,2\pi) $\\
\end{itemize}
Model $f_1(x)$ is close to the base model (almost a linear effect; this is the same model considered in \cite{klein-2015}), while $f_2(x)$ is a one cycle sinusoidal curve (highly non linear effect). In both scenarios, data are simulated as $y \sim N(f(\boldsymbol{x}), \tau_{\epsilon}^{-1})$, where covariate $\boldsymbol{x}=\{x_1,..,x_n\}$ takes values on a regular grid. We assume a standard P-spline model with one covariate, $y \sim N(\bm B \bm\beta, \tau_{\epsilon}^{-1})$, $\bm\beta \sim N(0, \tau_{\beta}^{-1} \bm R^{-1})$, where $\bm R$ is the structure of an IGMRF of order 2, and $\bm B$ contains $K$ cubic B-splines evaluated at $\bm x$. 

Different scenarios are generated by setting: $n=\{20,50\}$ (small and moderate sample sizes), $K=\{20,30\}$, $\tau_{\epsilon}=\{0.25,1,5\}$ (high, moderate and low noise). We aim to assess the model fit obtained by the following priors:  
\begin{itemize}
\item conjugate Gamma prior on $\tau_{\beta}$, with two specifications widely used in applications: \\
$\text{Gamma}(10^{-3},10^{-3})$ and $\text{Gamma}(1,5e-04)$.
\item our joint prior $\pi_{PC}(\tau_{\beta}|\tau_{\epsilon})\pi(\tau_{\epsilon})$, inducing a PC prior for degrees of freedom, $\pi_{PC}(d)$, with parameters $U$ and $\alpha$. We set $\alpha=0.01$ and various upper bounds $U=\{2,3,5,7,10\}$. Note that we specify $U=2$ only to check consistency of results in the limit case where any deviation from the base model is strongly penalized (in applications, this is not a sensible choice as it forces the fit towards a linear trend; for more details see the the joint prior in action with simulated data in the supplemental material). 
\item SD prior on $\tau_{\beta}$ \citep{klein-2015}, with $\alpha=0.01$ and three specifications for the scale of $f$, $c=\{1.5,2,3\}$. Note that, since both $f_1$ and $f_2$ vary within $(-1,1)$, $c=1.5$ seems the most sensible choice as an upper bound for the scale of both functions, resulting in a sufficiently flexible prior, while $c=\{2,3\}$ leads to an even more flexible prior. However, from Table \ref{tab-equivalent-U} we see that the degrees of freedom implied by an SD prior with parameter $c$ strongly depend on the noise present in the data (and to some extent on the adopted design): for instance, $c=1.5$ implies an upper bound for $d$ around $2.72$ in the high noise case (for the design $n=20$, $K=20$) which results into a very restrictive prior in fact.
\end{itemize}

\begin{table}
\small
\centering
\begin{tabular}{|l||l|l|l||l|l|l||l|l|l||}
\hline
 &\multicolumn{3}{l||}{High noise ($\tau_{\epsilon}=0.25$)}&\multicolumn{3}{l||}{Moderate noise ($\tau_{\epsilon}=1$)}&\multicolumn{3}{l||}{Low noise ($\tau_{\epsilon}=5$)}\\
\cline{2-10}
B-spline design & $c=1.5$& $c=2$ & $c=3$ & $c=1.5$ & $c=2$ & $c=3$ & $c=1.5$ & $c=2$ & $c=3$ \\
\hline\hline
$n=20$; $K=20$ & $2.72$ & $2.99$ & $3.41$ & $3.41$ & $3.78$ & $4.36$ & $4.54$ & $5.08$ & $ 5.89$\\
$n=50$; $K=20$ & $3.11$ & $3.44$ & $3.95$ & $3.95$ & $4.40$ & $5.10$ & $5.31$ & $5.95$ & $6.90$\\
$n=20$; $K=30$ & $2.70$ & $2.95$ & $3.44$ & $3.39$ & $3.75$ & $4.43$ & $4.54$ & $5.06$ & $6.04$ \\
$n=50$; $K=30$ & $3.09$ & $3.40$ & $4.00$ & $3.93$ & $4.37$ & $5.20$ & $5.34$ & $5.98$ & $ 7.16$ \\
\hline 
\end{tabular}
\caption{Implied degrees of freedom, $d$, for the SD prior. The entries in the table refer to the upper bound, $U$, for $d$, obtained by assuming an SD prior with parameters $c$ and $\alpha=0.01$, in the different simulation scenarios. The computation of $U$ involves the use of the \texttt{sdPrior} \texttt{R} package \citep{sdPrior}; for more details see the supplemental material.}
\label{tab-equivalent-U}
\end{table}

\subsection{Results}
In each scenario, goodness of fit was assessed for each of $1000$ simulated datasets by the mean squared error ($MSE$) of $\hat f(\bm x)=\bm B \hat{\bm\beta}$, where $\hat{\bm\beta}$ is the posterior mean, as $MSE=n^{-1}\sum_{i=1}^n (\hat f(x_i) - f(x_i))^2$. 
The posterior for $\bm \beta$ was computed using INLA \citep{rue-inla} for the Gamma and SD priors, and using MCMC algorithm \ref{mcmc1} for the joint prior. For the sake of comparison between the three classes of priors, we assume $\pi(\tau_{\epsilon})=\text{Gamma}(1,5e-04)$ throughout the simulation study (hence, our joint prior is $\pi_{PC}(\tau_{\beta}|\tau_{\epsilon})\text{Gamma}(1,5e-04)$). 

\begin{figure}
\centerline{
\includegraphics[scale=.7]{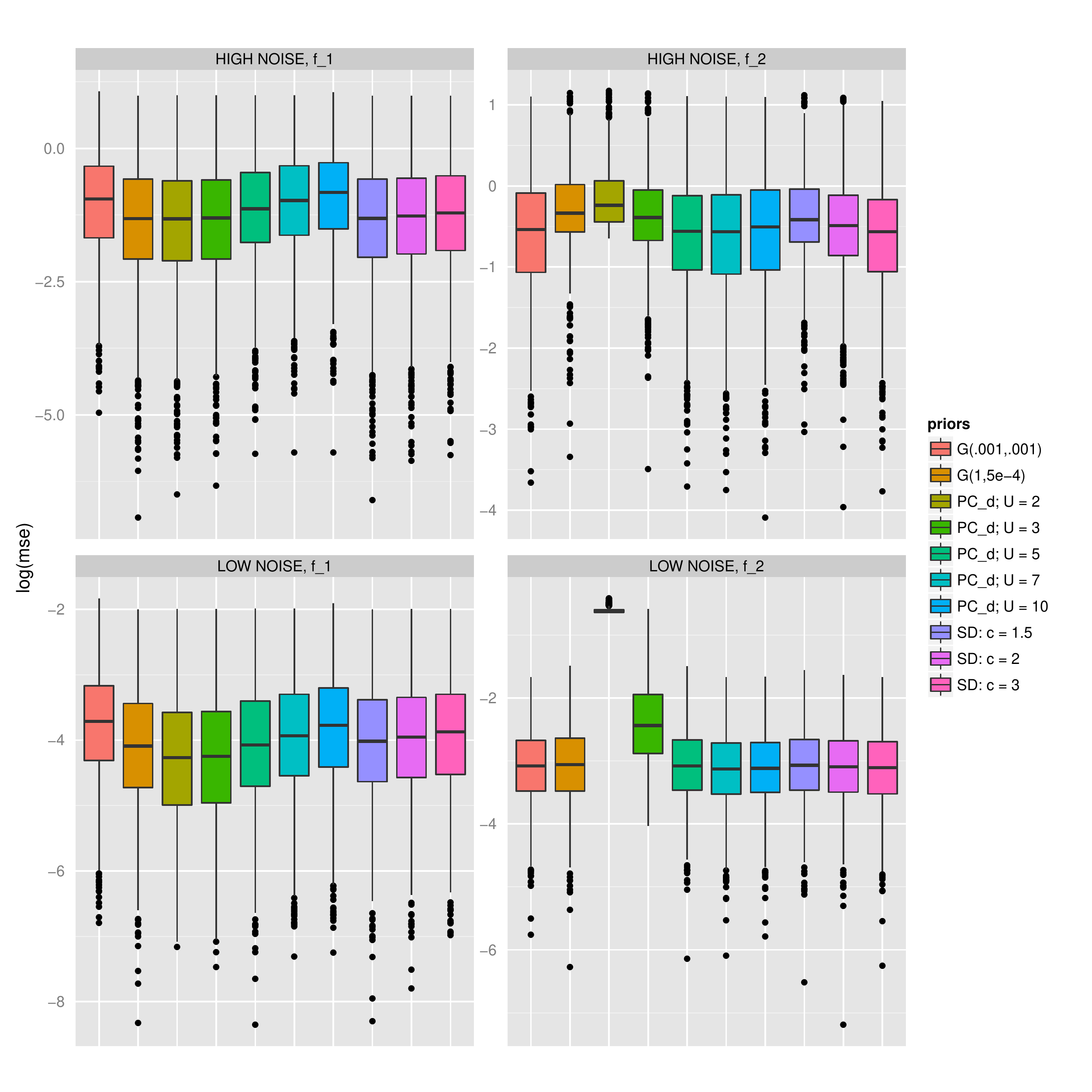}}
\caption{Simulation results: $\log(MSE)$ for $f_1$ (left panels) and $f_2$ (right panels), in presence of high noise ($\tau_{\epsilon}=0.25$, top panels) and low noise ($\tau_{\epsilon}=5$, bottom panels), sample size $n=20$, $K=20$. In the legend on the right, label ``G'' indicates the Gamma prior; ``\text{PC\_d}'' indicates our PC prior for degrees of freedom (joint prior), with $\alpha=0.01$ and $U=\{2,3,5,7,10\}$; ``\text{SD}'' denotes scale dependent prior with $\alpha=0.01$ and $c=\{1.5,2,3\}$.}
\label{fig:compare-f1f2_n20-K20-HIGH-LOW-noise}
\end{figure}

Figure \ref{fig:compare-f1f2_n20-K20-HIGH-LOW-noise} reports $\log(MSE)$ for small sample size $n=20$ (which is when the hyper-prior is expected to be most influential on the posterior) and $K=20$. We do not see much change for increasing $n$ and $K$, thus results for other scenarios are only reported in the supplemental material. Our main findings are twofold:
\begin{enumerate}
\item The Gamma priors are generally outperformed by the two PC priors (both for degrees of freedom and scale), unless the data are very informative about the true model (e.g. in model $f_2$ with low noise, see bottom-right panel in Figure \ref{fig:compare-f1f2_n20-K20-HIGH-LOW-noise}). As expected, the $\text{Gamma}(10^{-3}, 10^{-3})$ overfits when model $f_1$ is the true one (see left panels in Figure \ref{fig:compare-f1f2_n20-K20-HIGH-LOW-noise}), while the $\text{Gamma}(1, 5e-04)$ performs poorly in  scenario $f_2$ (especially with high noise, see top-right panel in Figure \ref{fig:compare-f1f2_n20-K20-HIGH-LOW-noise}).
\item For sensible choices of the upper bound $U$ the joint prior performs better than, or at least as good as, the SD prior. The main difference is noticed in scenario $f_2$ with high noise (top-right panel in Figure \ref{fig:compare-f1f2_n20-K20-HIGH-LOW-noise}) and scenario $f_1$ with low noise  (bottom-left panel in Figure \ref{fig:compare-f1f2_n20-K20-HIGH-LOW-noise}). In the former, setting the joint prior with $U=\{5,7,10\}$ outperforms most SD prior specifications; in particular, the SD prior with $c=\{1.5,2\}$ achieves poor performance because the implied upper bound for $d$ is too small (i.e. below $3$, see Table \ref{tab-equivalent-U}) for this case, where the true effect is highly non linear and the data are very noisy. The SD prior with $c=3$ gives similar performance because it implies a larger upper bound for degrees of freedom.
Instead, in scenario $f_1$ with low noise, SD priors are outperformed by the joint prior with $U=\{2,3\}$, because any choice $c=\{1.5,2,3\}$ implies an upper bound for $d$ clearly larger than needed (i.e. above $4$, see Table \ref{tab-equivalent-U}) in this case, where data are very informative and the true effect is close to the base model.
\end{enumerate}

In summary, the joint prior accounts for the level of noise present in the data and performs well in general, provided that the elicited $U$ is an appropriate upper bound for the degrees of freedom of the true $f$.
Therefore, we conclude that the number of effective degrees of freedom is a relevant quantity for building PC priors for smoothing Gaussian data, and potentially better than other transformations of $\tau_{\beta}$ that do not depend on $\tau_{\epsilon}$.

\section{Application}
\label{sec:realdata}

We demonstrate the use of the joint prior within an additive P-spline framework for modelling nitrate concentration in river \textit{Oglio}, Lombardia region, Italy. A total of $n=576$ observations of NO3$^-$ concentration were collected during 2010-2012 by taking one sample at each season (spring, summer, autumn and winter) in 48 gauging stations located along the river catchment. The response variable is $\log(\texttt{NO3}^-_{ij})$, measured at station $i=1,...,48$ and season $j=1,...,4$. Covariate $\texttt{stream}_{i}$ is the distance from each station $i$ to the \textit{Iseo} lake (i.e., the river source) measured in $km$ along the stream network; the first station in proximity to the lake is at $\texttt{stream} \approx 0$, while the last station downstream the river is at $\texttt{stream} \approx 150$. The goal is to understand \textit{river enrichment} in terms of nitrates, by studying the behaviour of $\log(\texttt{NO3}^-)$ as the stream distance increases. A substantial amount of information comes from previous studies, see \cite{Delconte2014, bartoli-2012} and references therein, suggesting that river enrichment in terms of nitrates may vary non linearly as the stream distance increases. Due to different characteristics in terms of groundwater interactions and irrigation practices, the river catchment can be divided into upstream, middle and downstream reach. Different processes are expected within the three reaches and between seasons, hence the enrichment curve may show different shapes in the three river segments and seasons. 

In order to investigate possible seasonal effects on river enrichment, we adopt the model: $\log(\texttt{NO3}^-_{ij}) = \mu + \gamma_j + f_j(\texttt{stream}_{i}) + \epsilon_{ij}$, $i=1,...,48$, $j=1,...,4$, where $\mu$ is the overall intercept, $\gamma_j$ is the season-specific intercept and $f_j(\texttt{stream}_{i})$ is the season-specific smooth function of the stream distance, modelled with a P-spline with joint prior on $(\tau_{\beta_j},\tau_{\epsilon})$ as described in section \ref{sec:joint-pcprior},
\begin{eqnarray}
f_j(\texttt{stream}) &=& \bm B_j \bm \beta_j \ \ \ \ \ \ \ \ \ \ \ \ \ \ j =1,...,4 \nonumber \\
\boldsymbol{\beta_j}| \tau_{\beta_j} & \sim & N(0, \tau_{\beta_j}^{-1} \boldsymbol{R}^{-1}) \nonumber \\
\tau_{\beta_j} | \tau_{\epsilon} & \sim & \text{Gumbel}(1/2, \theta_j(\tau_{\epsilon}))  \label{eq:mod-oglio-cond-prior}\\ 
\tau_{\epsilon} & \sim & \pi(\tau_{\epsilon}) \propto 1/\tau_{\epsilon}.
\nonumber
\end{eqnarray}
We assume an IGMRF of order 2 with precision $\tau_{\beta_j}$ on each $\bm \beta_j$. Based on our prior information we specify an upper bound $U=8$ for the PC prior (\ref{eq:mod-oglio-cond-prior}), for all $j=1,...,4$, assuming that each $f_j$ is much more flexible than linear (i.e. $d>2$) and assigning 2 additional degrees of freedom to each of the three river segments, to capture possibly different enrichment behaviours. We believe this prior is flexible enough to describe the possible smooth change between the upstream middle and downstream behaviours. We set $\alpha=0.01$, saying that it is $1\%$ likely that $f_j$ is more flexible than $8$ degrees of freedom, $j=1,...,4$.

We fit the model using the additive P-splines under linear constraint algorithm \ref{mcmc2} in Appendix \ref{app:A2}. To define the matrices $\bm X$ and $\bm B$ we need to create suitable dummy vectors of length $n$: $\texttt{dummy}_j$, $j=2,...,4$, taking value 1 when the observation is from season $j$ and 0 elsewhere (these are associated to the season-specific intercepts); $\texttt{stream}*\texttt{dummy}_j$, $j=1,...,4$, taking the actual stream distance when the observation is from season $j$ and 0 elsewhere (these are associated to the season-specific slopes). The $n \text{ x } 8$ fixed effect design matrix is $\bm X=[\bm 1, \bm D, \bm S]$, with $\bm D=[\texttt{dummy}_2,...,\texttt{dummy}_4]$ and $\bm S=[\texttt{stream}*\texttt{dummy}_1,...,\texttt{stream}*\texttt{dummy}_4]$.
Each basis $\bm B_j$ contains $K=30$ cubic B-splines, evaluated on the season-specific stream distances $\texttt{stream}*\texttt{dummy}_j$. The $n \text{ x } (4 K)$ B-spline design matrix is $\bm B=[\bm B_1, \bm B_2, \bm B_3, \bm B_4]$.
To separate the season-specific slopes and intercepts from the season-specific smooth variation captured by the B-splines, suitable linear constraints must be applied to each $f_j$, as discussed in section \ref{sec:joint-pcprior-identifiability}. During MCMC iterations, the PC prior for each precision $\tau_{\beta_j}$ needs to be re-scaled according to the current $\tau_{\epsilon}$, but this can be done at a negligible computational cost.

\begin{figure}
\centerline{
\includegraphics[width=0.5\textwidth]{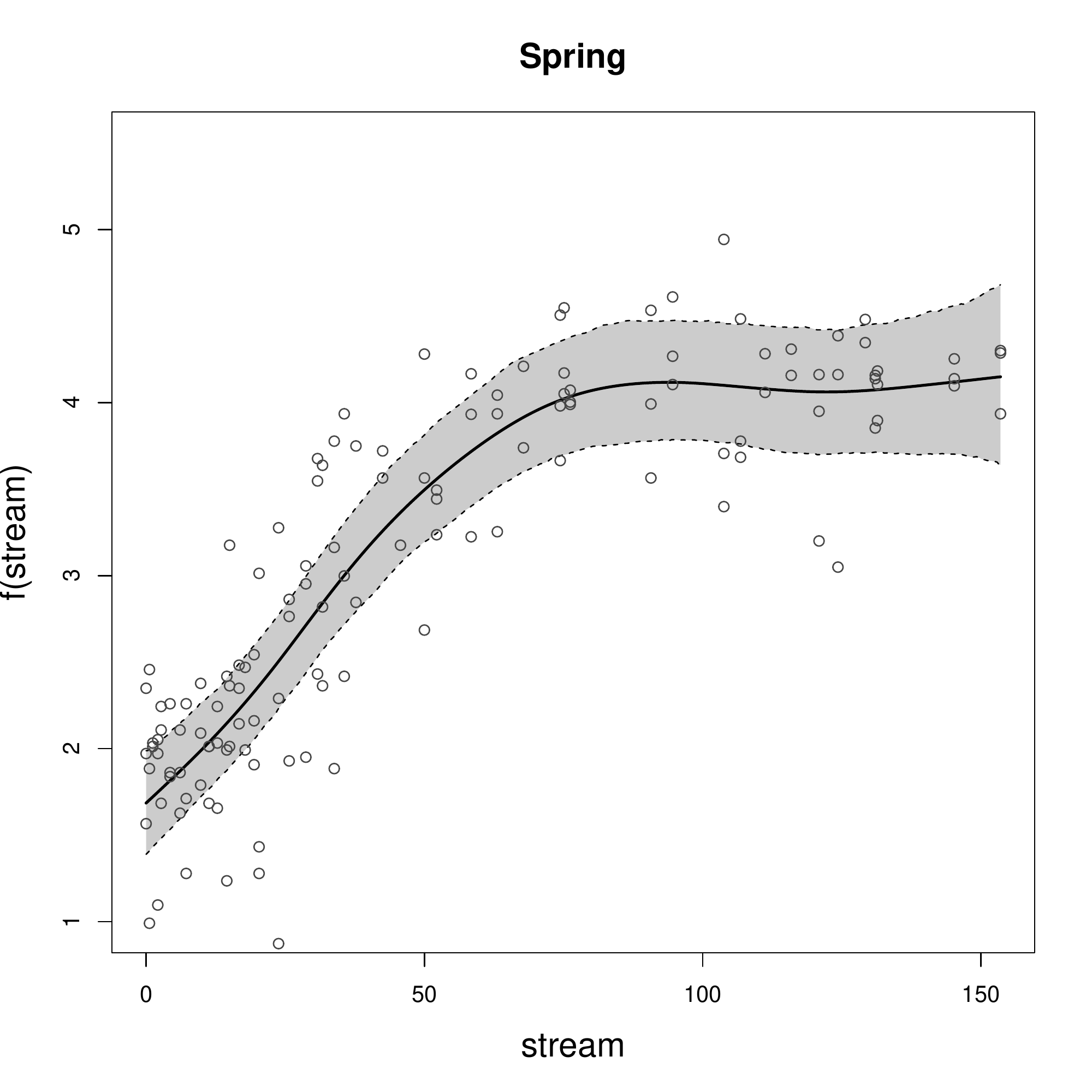}
\includegraphics[width=0.5\textwidth]{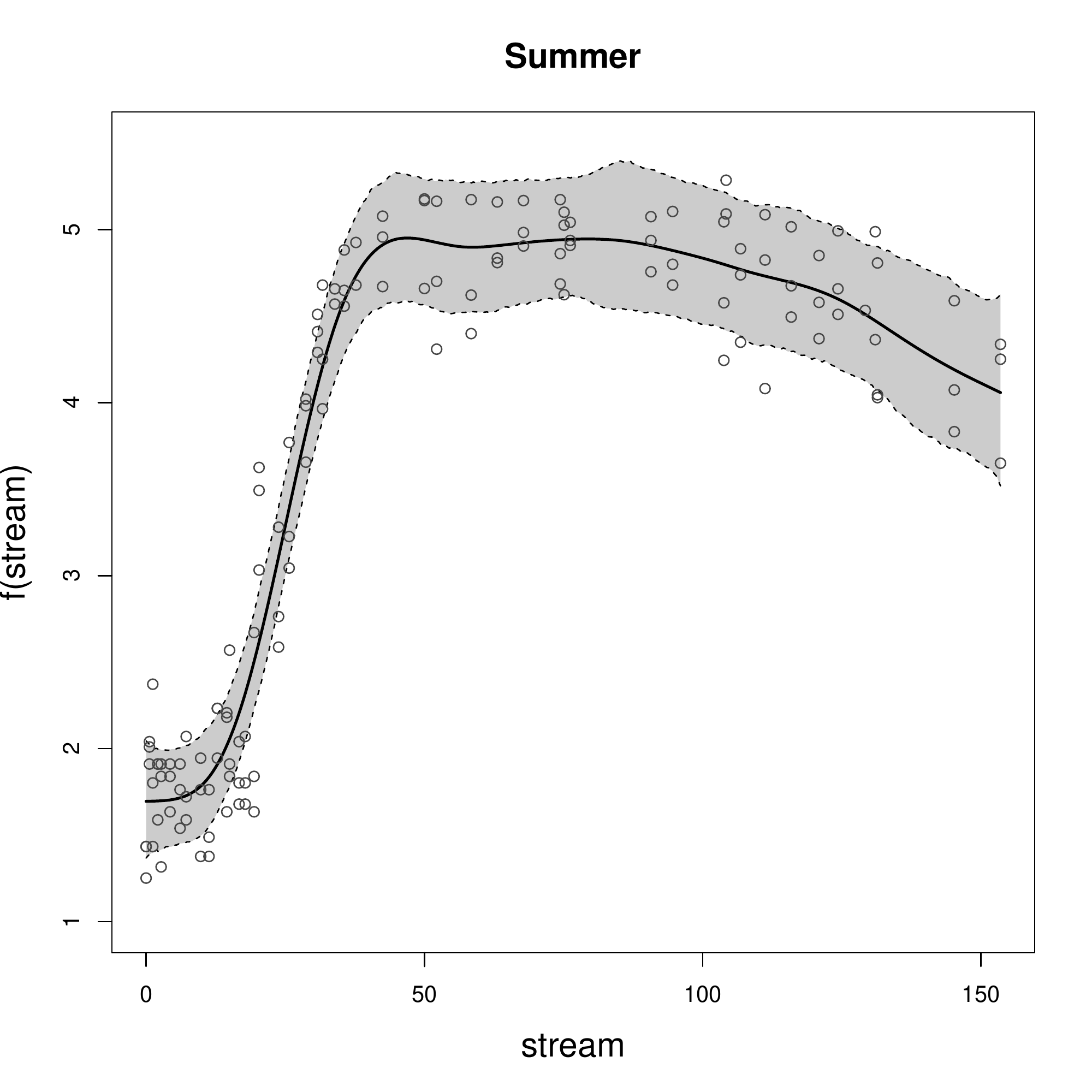}
}
\centerline{
\includegraphics[width=0.5\textwidth]{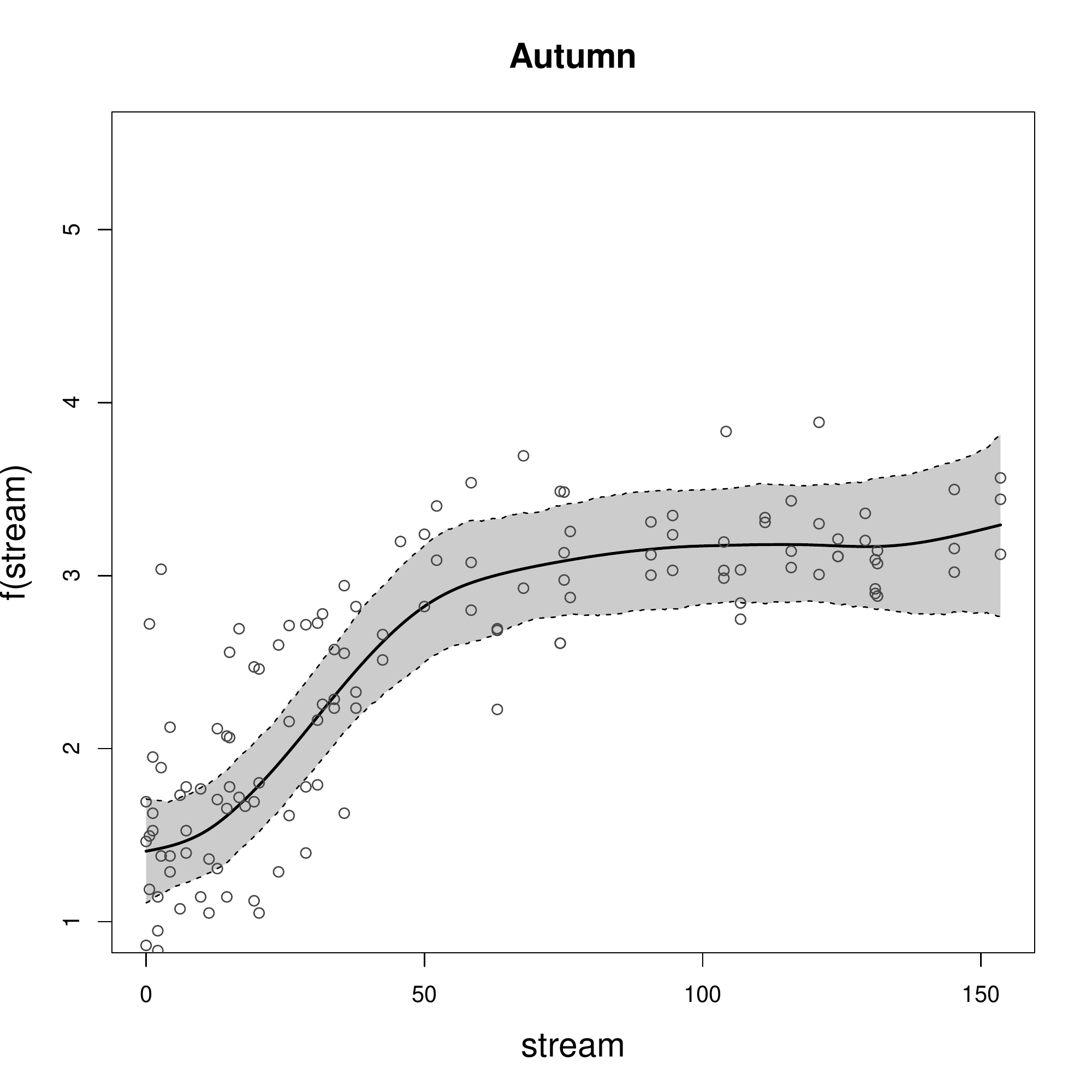}
\includegraphics[width=0.5\textwidth]{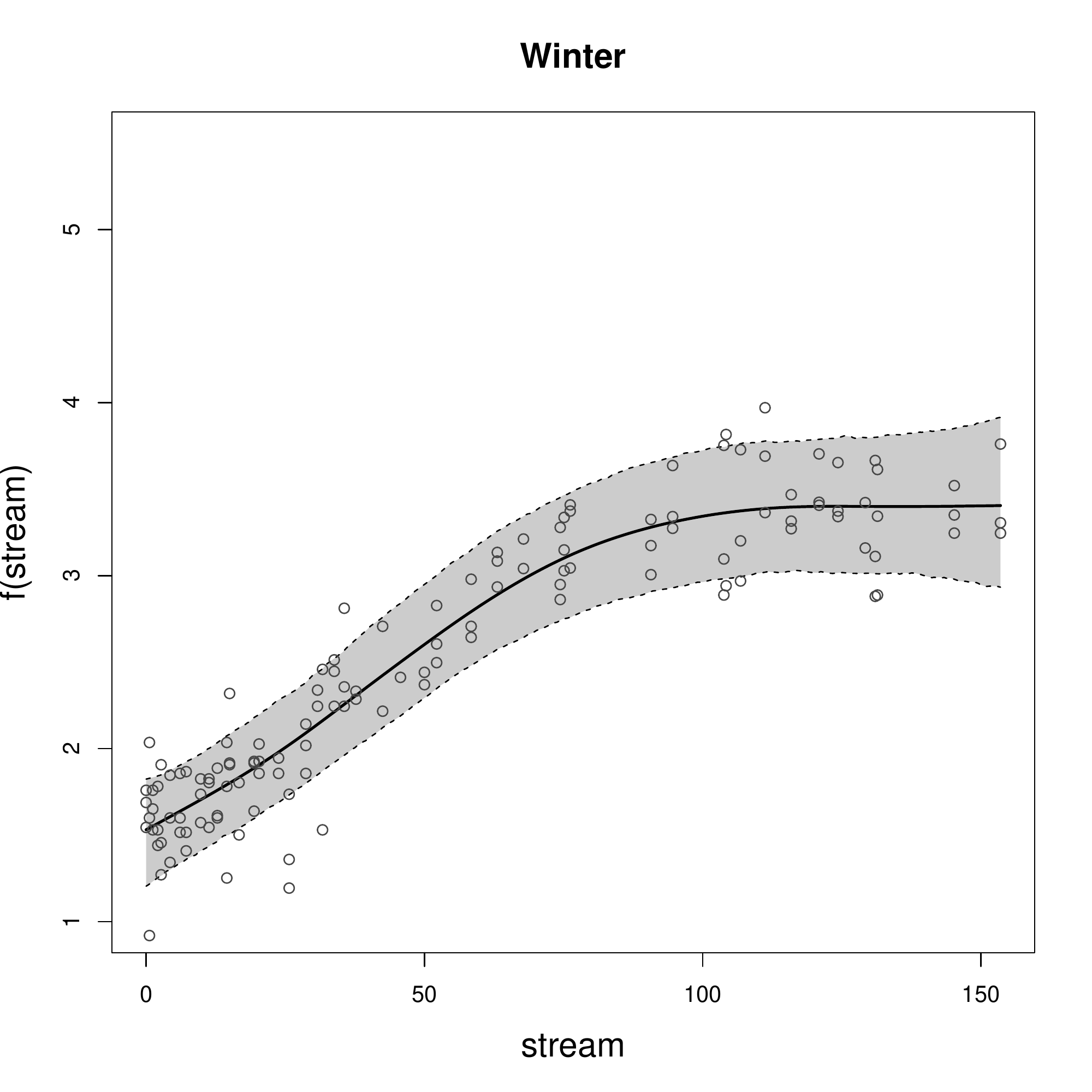}
}
\caption{Estimated river enrichment curves ($\hat f_j(\texttt{stream})$, black line), for each season, and $95\%$ credible bands (grey). The curve for summer shows clearly a distinctive pattern with respect to the other seasons. Partial residuals \citep{wood-2006} are plotted in each panel (dots), indicating larger variability, than what assumed by the model, for the log NO3$^{-}$ concentrations observed in spring and autumn (with possible outliers in the upstream river segment in autumn).}
\label{fig:oglio}
\end{figure}

The results displayed in Figure \ref{fig:oglio} reveal different river enrichment curves ($\hat f_j(\texttt{stream})$) in different seasons: a distinctive pattern is observed in summer, where the fitted curve shows a fast increase upstream ($\texttt{stream}< 50$ $km$) and tendency to decrease downstream the river ($\texttt{stream}> 80$ $km$). This pattern supports the argument given in \cite{Delconte2014} about an upstream reach (from 0 to 25 $km$) where nitrates are stable, reflecting the chemistry of the lake; a middle reach (from 25 to 80 $km$) showing an increase of NO3$^-$ concentration, probably due to groundwater inputs as replacement of river water abstracted for irrigation; a downstream reach (from 80 to 150 $km$) where nitrates should remain constant or even decrease, mainly due to the dilution of river water with NO3$^-$ deprived inputs.

\section{Discussion}
\label{sec:discussion}

PC priors are defined to penalize complexity with respect to a given base model, the magnitude of the penalty being elicited by the user using an intuitive  scaling approach. The scaling tool allows the user to derive the PC prior on an interpretable scale, different from the scale of the original parameter, provided that the link between the two is known. We took advantage of this nice feature and derived PC priors for the number of effective degrees of freedom $d$ of a P-spline model for Gaussian data.

For non Gaussian responses, the idea presented in this paper follows straightforwardly by assuming the definition of the degrees of freedom of a generalized P-spline model \citep{hastie-1990-book}, $d=\text{tr} \left(\boldsymbol{B}^{\textsf{T}}\bm W \boldsymbol{B} + \frac{\tau_{\beta}}{\tau_{\epsilon}}\boldsymbol{R}\right)^{-1} \bm B^{\textsf{T}}\bm W\bm B$, where $\bm W$ is a diagonal matrix, with entries depending on the linear predictor of the model (i.e. on $\bm B \bm \beta$) and the adopted link function. MCMC methods similar to those proposed in this paper can then be developed being careful about implementing the mapping $d(\tau_{\beta}|\tau_{\epsilon},\bm W)$, which is conditional on $\bm W$ in the generalized case (note that for most distributions in the exponential family $\tau_{\epsilon}$ is known, e.g. for Poisson we have $\tau_{\epsilon}=1$). For Poisson and Binomial responses an easier approach is to use auxiliary variable methods \citep{FruhwirthSchnatter-2007, FruhwirthSchnatter-2008} and work with an equivalent \textit{augmented} P-spline model for Gaussian (pseudo) data. The PC prior for $d$ can then be defined and implemented in the same way as described in this paper, with algorithms \ref{mcmc1} and \ref{mcmc2} needing only the inclusion of a Gibbs-step to update the augmented parameters.

The potential advantages of using PC priors for degrees of freedom are twofold. First, they are \textit{easy-to-elicit} by the user, who has to define two intuitive scaling parameters: $U$, an upper bound for $d$, and $\alpha$, the prior mass assigned to $d>U$. This scaling tool can be handled flexibly. For instance, elicitation of the median $M$ for the degrees of freedom results from fixing $\alpha=0.5$. In this case, the PC prior density is bimodal: one mode is set at the base model (by definition) and another mode is set around $M$ degrees of freedom. This bimodal behaviour is due to the attraction to the base model implicit in PC priors. 

As a second advantage, these PC priors avoid overfitting and are invariant under design, which means that the parameters $U$ and $\alpha$ do not need to be rescaled if the design changes. In other words, the PC prior is able to code into the model the prior knowledge on the complexity of the curve, or its degrees of freedom, in a design-adaptive way. 
The ability to adapt to design and avoid overfitting by construction, makes $\pi_{PC}(d)$ an appealing default choice in additive models where the latent structure includes several smooth functions (built on a basis of B-splines, e.g. P-splines) and other types of structures, such as individual random effects, spatial and spatio-temporal random effects.

\section*{Acknowledgements}
Massimo Ventrucci is funded by a FIRB 2012 grant (project nr. RBFR12URQJ, title: Statistical modeling of environmental phenomena: pollution, meteorology, health and their interactions), for research projects of national interest provided by the Italian Ministry of Education, Universities and Research. 
The dataset used in section 7 was kindly provided by the \textit{Consorzio dell'Oglio}, from the project ``Experimental assessment of the environmental flow in the lower Oglio river''. We thank Erica Racchetti and Alex Laini, Department of Life Science, University of Parma, for introducing us to the application in section 7 and for fruitful discussion on the ecological interpretation of results. Finally, we would like to thank the AE and two anonymous referees for their helpful comments.

\bibliographystyle{apalike}
\bibliography{biblio}

\begin{thebibliography}{}

\bibitem[Bartoli et~al., 2012]{bartoli-2012}
Bartoli, M., Racchetti, E., Delconte, C.~A., Sacchi, E., Soana, E., Laini, A.,
  Longhi, D., and Viaroli, P. (2012).
\newblock Nitrogen balance and fate in a heavily impacted watershed ({O}glio
  {R}iver, {N}orthern {I}taly): in quest of the missing sources and sinks.
\newblock {\em Biogeosciences}, 9(1):361--373.

\bibitem[Belitz et~al., 2000]{bayesx}
Belitz, C., Brezger, A., Kneib, T., Lang, S., and Umlauf, N. (2000).
\newblock Bayes{X} software for {B}ayesian inference in structured additive
  regression models.
\newblock Technical report.
\newblock Available from \url{http://www.stat.uni-muenchen.de/~bayesx}.

\bibitem[Currie et~al., 2006]{currie-2006}
Currie, I., Durb\'{a}n, M., and Eilers, P. (2006).
\newblock Generalized linear array models with applications to multidimensional
  smoothing.
\newblock {\em J. R. Statist. Soc. B}, 68:259--280.

\bibitem[Delconte et~al., 2014]{Delconte2014}
Delconte, C., Sacchi, E., Racchetti, E., Bartoli, M., Mas-Pla, J., and Re, V.
  (2014).
\newblock Nitrogen inputs to a river course in a heavily impacted watershed: A
  combined hydrochemical and isotopic evaluation ({O}glio {R}iver {B}asin, {N}
  {I}taly).
\newblock {\em Science of The Total Environment}, 466–467:924 -- 938.

\bibitem[Eilers et~al., 2006]{eilers-2006}
Eilers, P., Currie, I., and Durb\'{a}n, M. (2006).
\newblock Fast and compact smoothing on large multidimensional grids.
\newblock {\em Computational Statistics \& Data Analysis}, 5:61--76.

\bibitem[Eilers and Marx, 1996]{eilers-1996}
Eilers, P. and Marx, B. (1996).
\newblock Flexible smoothing with {B}-splines and penalties.
\newblock {\em Statistical Science}, 11:89--121.

\bibitem[Eilers and Marx, 2010]{eilers-2010}
Eilers, P. and Marx, B. (2010).
\newblock Splines, knots, and penalties.
\newblock {\em Wiley Interdisciplinary Reviews: Computational Statistics},
  2:637--653.

\bibitem[Fahrmeir and Kneib, 2009]{Fahrmeir_kneibt-2009}
Fahrmeir, L. and Kneib, T. (2009).
\newblock Propriety of posteriors in structured additive regression models:
  Theory and empirical evidence.
\newblock {\em Journal of Statistical Planning and Inference}, 139:843--859.

\bibitem[Fahrmeir et~al., 2004]{Fahrmeir04penalizedstructured}
Fahrmeir, L., Kneib, T., and Lang, S. (2004).
\newblock Penalized structured additive regression for space-time data: a
  {B}ayesian perspective.
\newblock {\em STATISTICA SINICA}, 14:715--745.

\bibitem[Fahrmeir et~al., 2013]{fahrmeir-2013-book}
Fahrmeir, L., Kneib, T., Lang, S., and Marx, B. (2013).
\newblock {\em Regression: models, methods and applications}.
\newblock Springer-Verlag, Berlin.

\bibitem[Fong et~al., 2010]{fong-2010}
Fong, Y., Rue, H., and Wakefield, J. (2010).
\newblock Bayesian inference for generalized linear mixed models.
\newblock {\em Biostatistics}, 11(3):397--412.

\bibitem[Frühwirth-Schnatter and Frühwirth, 2007]{FruhwirthSchnatter-2007}
Frühwirth-Schnatter, S. and Frühwirth, R. (2007).
\newblock Auxiliary mixture sampling with applications to logistic models.
\newblock {\em Computational Statistics \& Data Analysis}, 51(7):3509 -- 3528.

\bibitem[Fr{\"u}hwirth-Schnatter et~al., 2008]{FruhwirthSchnatter-2008}
Fr{\"u}hwirth-Schnatter, S., Fr{\"u}hwirth, R., Held, L., and Rue, H. (2008).
\newblock Improved auxiliary mixture sampling for hierarchical models of
  non-gaussian data.
\newblock {\em Statistics and Computing}, 19(4):479--492.

\bibitem[Fr{\"u}hwirth-Schnatter and Wagner, 2010]{FruhwirthSchnatter-2010}
Fr{\"u}hwirth-Schnatter, S. and Wagner, H. (2010).
\newblock Stochastic model specification search for {G}aussian and partial
  non-{G}aussian state space models.
\newblock {\em Journal of Econometrics}, 154(1):85 -- 100.

\bibitem[Fr{\"u}hwirth-Schnatter and Wagner, 2011]{FruhwirthSchnatter-2011}
Fr{\"u}hwirth-Schnatter, S. and Wagner, H. (2011).
\newblock Bayesian variable selection for random intercept modeling of
  {G}aussian and non-{G}aussian data.
\newblock In {\em J. M. Bernardo, M. J. Bayarri, J. O. Berger, A. P. Dawid, D.
  Heckerman, A. F. M. Smith and M. West (Eds.)}, pages 165--200. Bayesian
  Statistics 9, Oxford.

\bibitem[Gelman, 2006]{gelman-2006}
Gelman, A. (2006).
\newblock Prior distributions for variance parameters in hierarchical models
  (comment on article by {B}rowne and {D}raper).
\newblock {\em Bayesian Anal.}, 1(3):515--534.

\bibitem[Hastie and Tibshirani, 1990]{hastie-1990-book}
Hastie, T. and Tibshirani, R. (1990).
\newblock {\em Generalized Additive Models}.
\newblock Chapman and Hall, London.

\bibitem[Hastie et~al., 2009]{hastie-element-statistical-learning}
Hastie, T., Tibshirani, R., and Friedman, J. (2009).
\newblock {\em The Elements of Statistical Learning}.
\newblock Springer Series in Statistics. Springer-Verlag New York.

\bibitem[Jullion and Lambert, 2007]{Jullion-2007}
Jullion, A. and Lambert, P. (2007).
\newblock Robust specification of the roughness penalty prior distribution in
  spatially adaptive bayesian p-splines models.
\newblock {\em Computational Statistics \& Data Analysis}, 51(5):2542 -- 2558.

\bibitem[Klein, 2015]{sdPrior}
Klein, N. (2015).
\newblock {\em sdPrior: Scale-Dependent Hyperpriors in Structured Additive
  Distributional Regression}.
\newblock R package version 0.3.

\bibitem[Klein and Kneib, 2015]{klein-2015}
Klein, N. and Kneib, T. (2015).
\newblock Scale-dependent priors for variance parameters in structured additive
  distributional regression.
\newblock {\em Bayesian Analysis}.

\bibitem[Knorr-Held and Rue, 2002]{held-rue-2002}
Knorr-Held, L. and Rue, H. (2002).
\newblock On block updating in {M}arkov {r}andom {f}ield models for diasease
  mapping.
\newblock {\em Scandinavian Journal of Statistics}, 29(4):597--614.

\bibitem[Kullback and Leibler, 1951]{kld-1951}
Kullback, S. and Leibler, R.~A. (1951).
\newblock On information and sufficiency.
\newblock {\em The Annals of Mathematical Statistics}, 22:79--86.

\bibitem[Lang and Brezger, 2004]{brezger-2004}
Lang, S. and Brezger, A. (2004).
\newblock Bayesian {P}-splines.
\newblock {\em Journal of Computational and Graphical Statistics}, 13:183--212.

\bibitem[Rue and Held, 2005]{rue-2005}
Rue, H. and Held, L. (2005).
\newblock {\em Gaussian Markov Random Fields}.
\newblock Chapman and Hall/CRC.

\bibitem[Rue et~al., 2009]{rue-inla}
Rue, H., Martino, S., and Chopin, N. (2009).
\newblock Approximate bayesian inference for latent gaussian models using inte-
  grated nested laplace approximations (with discussion).
\newblock {\em Journal of the Royal Statistical Society, Series B},
  71(2):319--392.

\bibitem[Ruppert et~al., 2003]{ruppert2003semiparametric}
Ruppert, D., Wand, P., and Carroll, R. (2003).
\newblock {\em Semiparametric Regression}.
\newblock Cambridge Series in Statistical and Probabilistic Mathematics.
  Cambridge University Press.

\bibitem[{Simpson} et~al., 2014]{pcprior}
{Simpson}, D.~P., {Rue}, H., {Martins}, T.~G., {Riebler}, A., and {S{\o}rbye},
  S.~H. (2014).
\newblock {Penalising model component complexity: A principled, practical
  approach to constructing priors}.
\newblock {\em ArXiv e-prints}.

\bibitem[Sørbye and Rue, 2014]{sorbye-2013}
Sørbye, S.~H. and Rue, H. (2014).
\newblock Scaling intrinsic gaussian markov random field priors in spatial
  modelling.
\newblock {\em Spatial Statistics}, 8:39 -- 51.
\newblock Spatial Statistics Miami.

\bibitem[Wakefield, 2013]{wakefield-book}
Wakefield, J. (2013).
\newblock {\em Bayesian and Frequentist Regression Methods}.
\newblock Springer Series in Statistics. Springer-Verlag New York.

\bibitem[Wood, 2006]{wood-2006}
Wood, S. (2006).
\newblock {\em Generalized Additive Models: An Introduction with R}.
\newblock Chapman and Hall/CRC.

\end{thebibliography}

\newpage

\appendix
\section{Proofs}
\label{app:A1}

We show that $\pi_{PC}(d) \to \infty$ as $d \to r$, for $\theta>0$. To simplify notation in the following we write $\tau$ instead of $\tau_{\beta}$. Let us consider degrees of freedom $d$ expressed as a function of $\tau$, 
given a positive and finite $\tau_{\epsilon}$ and assuming $v_1'<v_2'<...< v_{K-r}'$ are the ordered positive eigenvalues of $\tau_{\epsilon}^{-1}\boldsymbol{R}(\boldsymbol{B}^{\textsf{T}}\boldsymbol{B})^{-1}$,
\begin{eqnarray}
d(\tau) & = &  r + \sum_{k=1}^{K-r} \frac{1}{1+ \tau v_k'} \nonumber \\
   & = & r + \frac{1}{1+ \tau v_1'} \cdot \left( 1+ \frac{1/\tau+  v_1'}{1/\tau+  v_2'} + ... + \frac{1/\tau+ v_1'}{1/\tau+ v_{K-r}'}\right) \label{constant} 
\end{eqnarray}
When $\tau \to \infty$, the term inside the bracket on the right hand side of equation (\ref{constant}) is a constant, then $d$ behaves like $ r + \frac{1}{1+ \tau v_1'}$. 
Since $d \to r$ if and only if $w \to 0$, $w=\frac{\sigma^2}{\sigma^2+v_1'}$, it is sufficient to study the behaviour of $\pi_{PC}(w)$ when $w \to 0$. 
By a change of variable, $\pi_{PC}(w) = \theta \exp\left(-\theta  \cdot const \sqrt{w}\right) \left| \frac{const}{2}\frac{1}{\sqrt{w}} \right|$, for positive $\sigma$ (using the fact that $\pi_{PC}(\sigma)$ is exponential with rate $\theta$; see \cite{pcprior}). When $w \to 0$, $\pi_{PC}(w) \propto  \frac{1}{\sqrt{w}}= \infty$, for $\theta>0$, which completes the proof. 

We now prove that $\pi_{G}(d)$ goes to $0$ as $d \to r$, for $a,b,>0$. Using the same argument as in the proof above, it is sufficient to check that, for positive $w$, $\pi_{G}(w) \propto w^{-a-1} \exp(-b/w) \to 0$ as $w \to 0$, because $\lim_{w \to 0^{+}} \exp(-b/w) = 0$. 

\newpage

\section{MCMC}
\label{app:A2}

\begin{algorithm}
\caption{MCMC for fitting a P-spline model $\bm y = \bm B \bm \beta + \bm \epsilon$, assuming the joint prior $\pi_{PC}(\tau_{\beta}|\tau_{\epsilon})\pi(\tau_{\epsilon})$. (In step \ref{algo1:canonical}, $N_C$ indicates the canonical parametrization of a GMRF, see \cite{rue-2005} chapter 2.2.)}
\label{mcmc1}
\begin{algorithmic}[1]
\STATE Initialise $\tau_{\epsilon}^{(0)}$, $\tau_{\beta}^{(0)}$, $\bm\beta^{(0)}$ 
\FOR{$j=1...,\text{n.iter}$}
\STATE sample $\tau_{\epsilon}^{*} \sim q(\tau_{\epsilon}^{*}|\tau_{\epsilon}^{(j-1)})$
\STATE sample $\tau_{\beta}^{*} \sim q(\tau_{\beta}^{*}|\tau_{\beta}^{(j-1)})$
\STATE sample $\bm\beta^{*} \sim N_{C}(\tau_{\epsilon}^{*} \bm{B}^{\textsf{T}}\bm{y}, \tau_{\epsilon}^{*} \bm{B}^{\textsf{T}}\bm{B} + \tau_{\beta}^{*} \bm{R})$
\label{algo1:canonical}
\STATE Rescale $\pi_{PC}(\tau_{\beta}|\tau_{\epsilon})$, according to the proposed $\tau_{\epsilon}^{*}$: \\
\begin{itemize}
\item evaluate the inverse mapping $d^{-1}(U|\tau_{\epsilon}^{*})$ and compute $\theta(\tau_{\epsilon}^{*})$, \\ the rescaled PC prior is a $\text{Gumbel}(1/2, \theta(\tau_{\epsilon}^{*}))$
\end{itemize}
\STATE sample $r \sim U(0,1)$
\IF {$r < \text{min}\left(1, \frac{\pi(\tau_{\epsilon}^{*}, \tau_{\beta}^{*}| \bm{y})}{\pi(\tau_{\epsilon}^{(j-1)}, \tau_{\beta}^{(j-1)}| \bm{y})} \right)$} \label{algo1:accept}
\STATE $\tau_{\epsilon}^{(j)} \leftarrow \tau_{\epsilon}^{*}$
\STATE $\tau_{\beta}^{(j)} \leftarrow \tau_{\beta}^{*}$
\STATE $\bm{\beta}^{(j)} \leftarrow \bm{\beta}^{*}$
\STATE $\theta(\tau_{\epsilon}^{(j)}) \leftarrow \theta(\tau_{\epsilon}^{*})$
\ENDIF 
\ENDFOR
\end{algorithmic}
\end{algorithm}

\begin{algorithm}
\caption{MCMC for fitting an additive P-spline model under linear constraints, $\bm y = \bm X \bm \gamma + \bm B \bm \beta^{ULC}+ \bm\epsilon$, assuming the joint prior $\pi_{PC}(\tau_{\beta}|\tau_{\epsilon})\pi(\tau_{\epsilon})$. The pseudo-code below considers only one covariate $\bm x$ (i.e., using notation of section \ref{sec:joint-pcprior-identifiability}, $J=1$ and $\bm X=\left[\bm 1, \bm x\right]$; extension to $J>1$ covariates is straightforward, by block updating $(\tau_{\beta_j}^*, \bm \beta_j^{*ULC})$, for each $j=1,...,J$)}
\label{mcmc2}
\begin{algorithmic}[1]
\STATE define $2 \text{ x } K$ matrix $\bm A=\left[
(\bm c^{\textsf{T}} \bm B)^{\textsf{T}}, 
(l^{\textsf{T}} \bm B)^{\textsf{T}}  
\right]^{\textsf{T}}$ with 2 linear constraints (note: in our experience it is convenient to rescale $\bm A$, dividing each row by its maximum)
\STATE Initialise $\tau_{\epsilon}^{(0)}$, $\tau_{\beta}^{(0)}$, $\bm\gamma^{(0)}$, $\bm\beta^{(0)}$; $\tau_{\gamma}$ is fixed
\FOR{$j=1...,\text{n.iter}$}
\STATE sample $\tau_{\epsilon}^{*} \sim q(\tau_{\epsilon}^{*}|\tau_{\epsilon}^{(j-1)})$
\STATE sample $\bm \gamma^{*} \sim N_C(\tau_{\epsilon}^{*} \bm{X}^{\textsf{T}}\bm{\tilde y}, \tau_{\epsilon}^{*} \bm{X}^{\textsf{T}}\bm{X} + \tau_{\gamma} \bm{I}_{J+1})$, with $\bm{\tilde y}=\bm y-\bm B\bm \beta^{ULC(j-1)}$
\STATE sample $r \sim U(0,1)$
\IF {$r < \text{min}\left(1, \frac{\pi(\tau_{\epsilon}^{*}| \bm{y})}{\pi(\tau_{\epsilon}^{(j-1)}| \bm{y})} \right)$}
\STATE $\tau_{\epsilon}^{(j)} \leftarrow \tau_{\epsilon}^{*}$
\STATE $\bm{\gamma}^{(j)} \leftarrow \bm{\gamma}^{*}$
\STATE Rescale $\pi_{PC}(\tau_{\beta}|\tau_{\epsilon})$, according to the current $\theta(\tau_{\epsilon}^{(j)})$: \\
\begin{itemize}
\item evaluate the inverse mapping $d^{-1}(U|\theta(\tau_{\epsilon}^{(j)})$ and compute $\theta(\tau_{\epsilon}^{(j)})$, \\ the rescaled conditional PC prior is a $\text{Gumbel}(0.5, \theta(\tau_{\epsilon}^{(j)}))$
\end{itemize}
\ENDIF 
\STATE sample $\tau_{\beta}^{*} \sim q(\tau_{\beta}^{*}|\tau_{\beta}^{(j-1)})$
\STATE sample $\bm\beta^{*}|\bm A\bm\beta=\bm 0$, using algorithm 2.6 in \cite{rue-2005}, steps are:
\begin{itemize}
\item sample $\bm\beta^{*} \sim N_{C}(\tau_{\epsilon}^{(j)} \bm{B}^{\textsf{T}}\bm{\tilde y}, \tau_{\epsilon}^{(j)} \bm{B}^{\textsf{T}}\bm{B} + \tau_{\beta}^{*} \bm{R})$, with $\bm{\tilde y}=\bm y-\bm X\bm \gamma^{(j-1)}$
\item $\bm\beta^{*ULC} = \bm\beta^{*} - \bm Q^{-1} \bm A^{\textsf{T}} (\bm A \bm Q^{-1} \bm A^{\textsf{T}})^{-1} \bm A \bm\beta^{*}$, with $\bm Q= \tau_{\epsilon}^{(j)} \bm{B}^{\textsf{T}}\bm{B} + \tau_{\beta}^{*} \bm{R}$
\end{itemize}
\IF {$r < \text{min}\left(1, \frac{\pi(\tau_{\beta}^{*}| \bm{y})}{\pi( \tau_{\beta}^{(j-1)}| \bm{y})} \right)$} \label{algo2:accept-beta}
\STATE $\tau_{\beta}^{(j)} \leftarrow \tau_{\beta}^{*}$
\STATE $\bm{\beta}^{ULC(j)} \leftarrow \bm{\beta}^{*ULC}$
\ENDIF 
\ENDFOR
\end{algorithmic}
\end{algorithm}

\end{document}